%%%%%%%%%%%%%%%%%%%%%%%%  header.tex :  macro for tex commands

\newbox\SlashedBox 
\def\slashed#1{\setbox\SlashedBox=\hbox{#1}
\hbox to 0pt{\hbox to 1\wd\SlashedBox{\hfil/\hfil}\hss}{#1}}
\def\hboxtosizeof#1#2{\setbox\SlashedBox=\hbox{#1}
\hbox to 1\wd\SlashedBox{#2}}

% The following is necessary so that we can get a partial slash
% inside a math display... sigh.
\def\mathslashed#1{\setbox\SlashedBox=\hbox{$#1$}
\hbox to 0pt{\hbox to 1\wd\SlashedBox{\hfil/\hfil}\hss}#1}

\def\ifsmall{\iffalse}  % default is unreduced.
\def\titlepagefont{}  % default is ordinary font.

% the ps: landscape must be the first special command in order
% to get the first page in landscape mode -- so we go through some
% contortions to define TeXgraphics in the default case.
\def\DefineTeXgraphics{%
\special{ps::[global] /TeXgraphics { } def}}  % No need to do anything

\def\today{\ifcase\month\or January\or February\or March\or April\or May
\or June\or July\or August\or September\or October\or November\or
December\fi\space\number\day, \number\year}
\def\eatPrefix19{}
\def\Year{\expandafter\eatPrefix\the\year}
\newcount\hours \newcount\minutes
\def\monthname{\ifcase\month\or
January\or February\or March\or April\or May\or June\or July\or
August\or September\or October\or November\or December\fi}
\def\shortmonthname{\ifcase\month\or
Jan\or Feb\or Mar\or Apr\or May\or Jun\or Jul\or
Aug\or Sep\or Oct\or Nov\or Dec\fi}

\def\TimeStamp{\hours\the\time\divide\hours by60%
\minutes -\the\time\divide\minutes by60\multiply\minutes by60%
\advance\minutes by\the\time%
${\rm \shortmonthname}\cdot\if\day<10{}0\fi\the\day\cdot\the\year%
\qquad\the\hours:\if\minutes<10{}0\fi\the\minutes$}

%\DefineTeXgraphics}

%\DefineTeXgraphics}

%\DefineTeXgraphics}

%\DefineTeXgraphics}
 
% restores pagenumbers

%\def\draft{\centerline{\it Preliminary Draft}\vskip 0.4in}

\newif\ifdraftmode
\newif\ifleftlabels  % Labels in left margins as well, for European-size paper
\def\draftnote#1{{\it #1}}
% Stolen from harvmac.tex 04/08/92
%       use \nolabels to get rid of eqn, ref, and fig labels in draft mode
\def\nolabels{\def\wrlabeL##1{}\def\eqlabeL##1{}\def\reflabeL##1{}}
\def\writelabels{\def\wrlabeL##1{\leavevmode\vadjust{\rlap{\smash%
{\line{{\escapechar=` \hfill\rlap{\sevenrm\hskip.03in\string##1}}}}}}}%
\def\eqlabeL##1{{\escapechar-1\rlap{\sevenrm\hskip.05in\string##1}}}%
\def\reflabeL##1{\noexpand\rlap{\noexpand\sevenrm[\string##1]}}}
\def\writeleftlabels{\def\wrlabeL##1{\leavevmode\vadjust{\rlap{\smash%
{\line{{\escapechar=` \hfill\rlap{\sevenrm\hskip.03in\string##1}}}}}}}%
\def\eqlabeL##1{{\escapechar-1%
\rlap{\sixrm\hskip.05in\string##1}%
\llap{\sevenrm\string##1\hskip.03in\hbox to \hsize{}}}}%
\def\reflabeL##1{\noexpand\rlap{\noexpand\sevenrm[\string##1]}}}
\nolabels

\newdimen\fullhsize
\newdimen\hstitle
\hstitle=\hsize % default
\newdimen\hsbody
\hsbody=\hsize % default
\newdimen\hbodyoffset
\hbodyoffset=\hoffset % default
\newbox\leftpage
\def\abstract#1{#1}
\def\rotated{\special{ps: landscape}
\magnification=1200  % This line must come before we change vsize,
                     % since \magnification sets it to a fixed value.
\baselineskip=14pt
\global\hstitle=9truein\global\hsbody=4.75truein
\global\vsize=7truein\global\voffset=-.31truein
\global\hoffset=-0.54in\global\hbodyoffset=-.54truein
\global\fullhsize=10truein
\def\DefineTeXgraphics{%
\special{ps::[global] 
/TeXgraphics {currentpoint translate 0.7 0.7 scale
              -80 0.72 mul -1000 0.72 mul translate} def}}
 % 0.7 is slightly less than the ratio of horizontal sizes: 4.75 to 6.5
\let\lr=L
\def\ifsmall{\iftrue}
\def\titlepagefont{\twelvepoint}
\trueseventeenpoint
\def\almostshipout##1{\if L\lr \count1=1
      \global\setbox\leftpage=##1 \global\let\lr=R
   \else \count1=2
      \shipout\vbox{\hbox to\fullhsize{\box\leftpage\hfil##1}}
      \global\let\lr=L\fi}

\output={\ifnum\count0=1 %%% This is the HUTP version
 \shipout\vbox{\hbox to \fullhsize{\hfill\pagebody\hfill}}\advancepageno
 \else
 \almostshipout{\leftline{\vbox{\pagebody\makefootline}}}\advancepageno 
 \fi}

\def\abstract##1{{\leftskip=1.5in\rightskip=1.5in ##1\par}} }

% Messages on lines by themselves
\def\linemessage#1{\immediate\write16{#1}}

% tagged sec numbers
\global\newcount\secno \global\secno=0
\global\newcount\appno \global\appno=0
\global\newcount\meqno \global\meqno=1
\global\newcount\subsecno \global\subsecno=0
% and figure numbers
\global\newcount\figno \global\figno=0

\newif\ifAnyCounterChanged
% If we are comparing numbers, there's no special problem.
% But if we are comparing roman numerals, we must be careful, because
% stuff read in from the aux file would be made up of ordinary
% characters (category code = 11), whereas \romannumeral generates
% characters with category code = 12..., so the stuff from the
% current run won't appear equal to the previous definition, as far
% as \warnIfChanged is concerned.
% To get around this, we have a macro \makeNormal, which converts
% letters `ivxlcdmIVXLCDM' to normal letters, no matter what their category
% code.  The macro has the convoluted form it does, with aftergroup's & all,
% to avoid blowing up TeX...
% The macro is used below in makeNormalizedRomappno, by which means we
% define the appendix counters to be strings containing vanilla versions
% of the letters... Sigh
\let\terminator=\relax
% The string to be normalized must not contain { and } tokens...
\def\normalize#1{\ifx#1\terminator\let\next=\relax\else%
\if#1i\aftergroup i\else\if#1v\aftergroup v\else\if#1x\aftergroup x%
\else\if#1l\aftergroup l\else\if#1c\aftergroup c\else%
\if#1m\aftergroup m\else%
\if#1I\aftergroup I\else\if#1V\aftergroup V\else\if#1X\aftergroup X%
\else\if#1L\aftergroup L\else\if#1C\aftergroup C\else%
\if#1M\aftergroup M\else\aftergroup#1\fi\fi\fi\fi\fi\fi\fi\fi\fi\fi\fi\fi%
\let\next=\normalize\fi%
\next}
% makes #1 a normalized version of #2...
\def\makeNormal#1#2{\def\doNormalDef{\edef#1}\begingroup%
\aftergroup\doNormalDef\aftergroup{\normalize#2\terminator\aftergroup}%
\endgroup}
% makes a normalized version of its argument:

\def\warnIfChanged#1#2{%
\ifundef#1% skip it
\else\begingroup%
\edef\oldDefinitionOfCounter{#1}\edef\newDefinitionOfCounter{#2}%
%\message{old: \oldDefinitionOfCounter}%
%\message{new: \newDefinitionOfCounter}%
\ifx\oldDefinitionOfCounter\newDefinitionOfCounter%
\else%
\linemessage{Warning: definition of \noexpand#1 has changed.}%
\global\AnyCounterChangedtrue\fi\endgroup\fi}

\def\Section#1{\global\advance\secno by1\relax\global\meqno=1%
\global\subsecno=0%
\bigbreak\bigskip% (combination \goodbreak\bigskip\bigskip)
\centerline{\twelvepoint \bf %
\the\secno. #1}%
\par\nobreak\medskip\nobreak}
\def\tagsection#1{%
\warnIfChanged#1{\the\secno}%
\xdef#1{\the\secno}%
\ifWritingAuxFile\immediate\write\auxfile{\noexpand\xdef\noexpand#1{#1}}\fi%
}
\def\section{\Section}
\def\Subsection#1{\global\advance\subsecno by1\relax\medskip %
\leftline{\bf\the\secno.\the\subsecno\ #1}%
\par\nobreak\smallskip\nobreak}
\def\tagsubsection#1{%
\warnIfChanged#1{\the\secno.\the\subsecno}%
\xdef#1{\the\secno.\the\subsecno}%
\ifWritingAuxFile\immediate\write\auxfile{\noexpand\xdef\noexpand#1{#1}}\fi%
}

\def\subsection{\Subsection}

\def\romappno{\uppercase\expandafter{\romannumeral\appno}}
\def\makeNormalizedRomappno{%
\expandafter\makeNormal\expandafter\normalizedromappno%
\expandafter{\romannumeral\appno}%
\edef\normalizedromappno{\uppercase{\normalizedromappno}}}
\def\Appendix#1{\global\advance\appno by1\relax\global\meqno=1\global\secno=0%
\global\subsecno=0%
\bigbreak\bigskip% (combination \goodbreak\bigskip\bigskip)
\centerline{\twelvepoint \bf Appendix %
\romappno. #1}%
\par\nobreak\medskip\nobreak}
\def\tagappendix#1{\makeNormalizedRomappno%
\warnIfChanged#1{\normalizedromappno}%
\xdef#1{\normalizedromappno}%
\ifWritingAuxFile\immediate\write\auxfile{\noexpand\xdef\noexpand#1{#1}}\fi%
}
\def\appendix{\Appendix}
\def\Subappendix#1{\global\advance\subsecno by1\relax\medskip %
\leftline{\bf\romappno.\the\subsecno\ #1}%
\par\nobreak\smallskip\nobreak}
\def\tagsubappendix#1{\makeNormalizedRomappno%
\warnIfChanged#1{\normalizedromappno.\the\subsecno}%
\xdef#1{\normalizedromappno.\the\subsecno}%
\ifWritingAuxFile\immediate\write\auxfile{\noexpand\xdef\noexpand#1{#1}}\fi%
}

\def\eqn#1{\makeNormalizedRomappno%
\ifnum\secno>0%
  \warnIfChanged#1{\the\secno.\the\meqno}%
  \eqno(\the\secno.\the\meqno)\xdef#1{\the\secno.\the\meqno}%
     \global\advance\meqno by1
\else\ifnum\appno>0%
  \warnIfChanged#1{\normalizedromappno.\the\meqno}%
  \eqno({\rm\romappno}.\the\meqno)%
      \xdef#1{\normalizedromappno.\the\meqno}%
     \global\advance\meqno by1
\else%
  \warnIfChanged#1{\the\meqno}%
  \eqno(\the\meqno)\xdef#1{\the\meqno}%
     \global\advance\meqno by1
\fi\fi%
\eqlabeL#1%
\ifWritingAuxFile\immediate\write\auxfile{\noexpand\xdef\noexpand#1{#1}}\fi%
}
\def\defeqn#1{\makeNormalizedRomappno%
\ifnum\secno>0%
  \warnIfChanged#1{\the\secno.\the\meqno}%
  \xdef#1{\the\secno.\the\meqno}%
     \global\advance\meqno by1
\else\ifnum\appno>0%
  \warnIfChanged#1{\normalizedromappno.\the\meqno}%
  \xdef#1{\normalizedromappno.\the\meqno}%
     \global\advance\meqno by1
\else%
  \warnIfChanged#1{\the\meqno}%
  \xdef#1{\the\meqno}%
     \global\advance\meqno by1
\fi\fi%
\eqlabeL#1%
\ifWritingAuxFile\immediate\write\auxfile{\noexpand\xdef\noexpand#1{#1}}\fi%
}
\def\anoneqn{\makeNormalizedRomappno%
\ifnum\secno>0
  \eqno(\the\secno.\the\meqno)%
     \global\advance\meqno by1
\else\ifnum\appno>0
  \eqno({\rm\normalizedromappno}.\the\meqno)%
     \global\advance\meqno by1
\else
  \eqno(\the\meqno)%
     \global\advance\meqno by1
\fi\fi%
}
\def\mfig#1#2{\ifx#20%unnumbered figure
\else\global\advance\figno by1%
\relax#1\the\figno%
\warnIfChanged#2{\the\figno}%
\xdef#2{\the\figno}%
\reflabeL#2%
\ifWritingAuxFile\immediate\write\auxfile{\noexpand\xdef\noexpand#2{#2}}\fi\fi%
}

\catcode`@=11 % borrow the private macros of PLAIN (with care)

% \LoadFigure is used to put a figure into the text.  Its first argument
% is the symbolic name for the figure (if it isn't defined, a new number
% will be assigned);  the second argument is a caption;
% the third argument size information in the form 
% \epsfxsize=3.0in\epsfysize=3.5in (this argument may be blank and
% may contain any valid preparatory argument used by the epsf package);
% the fourth and last argument is the name of the file which contains the 
% figure.
% The macro is basically just a front-end for \epsfbox; its purpose is 
% to allow figures to be switched from placement in the running text
% to placement on a separate page at the end of the text.  This choice
% is made using the flag \FiguresInText{true,false}; in the latter case,
% figures are placed at the end, size information is ignored (figures
% will be full-size), and the captions are listed separately on a page
% when the \listfigs command is invoked, followed by the figures, each
% on a separate page.  
%  The epsf package must be loaded by the user.
%  To change the size of captions in the text, redefine \captionsize.
\newif\ifFiguresInText\FiguresInTexttrue
\newif\if@FigureFileCreated
\newwrite\capfile
\newwrite\figfile

%default
\newif\ifcaption
\captiontrue
\def\captionsize{\tenrm}
\def\PlaceTextFigure#1#2#3#4{%
\vskip 0.5truein%
#3\hfil\epsfbox{#4}\hfil\break%
\ifcaption\hfil\vbox{\captionsize Figure #1. #2}\hfil\fi%
\vskip10pt}
\def\PlaceEndFigure#1#2{%
\epsfxsize=\hsize\epsfbox{#2}\vfill\centerline{Figure #1.}\eject}

\def\LoadFigure#1#2#3#4{%
\ifundef#1{\phantom{\mfig{}#1}}\else%  Write out definition only if it's new.
%\ifx#10% unnumbered figure
%\else\warnIfChanged#1{\the\figno}%
%\ifWritingAuxFile\immediate\write\auxfile{\noexpand\xdef\noexpand#1{#1}}\fi\fi
\fi%
\ifFiguresInText% Figure is immediate
\PlaceTextFigure{#1}{#2}{#3}{#4}%
\else% Figure is at the end
\if@FigureFileCreated\else%
\immediate\openout\capfile=\jobname.caps%
\immediate\openout\figfile=\jobname.figs%
@FigureFileCreatedtrue\fi%
\immediate\write\capfile{\noexpand\item{Figure \noexpand#1.\ }{#2}\vskip10pt}%
\immediate\write\figfile{\noexpand\PlaceEndFigure\noexpand#1{\noexpand#4}}%
\fi}

\def\listfigs{\ifFiguresInText\else%
\vfill\eject\immediate\closeout\capfile%\parindent=20pt
\immediate\closeout\figfile%
\centerline{{\bf Figures}}\bigskip\frenchspacing%
\catcode`@=11 % borrow the private macros of PLAIN (with care)
\def\captionsize{\tenrm}
\input \jobname.caps\vfill\eject\nonfrenchspacing%
\catcode`\@=\active
\catcode`@=12  % No longer.
\input\jobname.figs\fi}

%\font\titlefont=cmr10 at 16pt
\font\ninerm=cmr9
\font\eightrm=cmr8
\font\sixrm=cmr6

\def\loadtrueseventeenpoint{
 \font\seventeenrm=cmr10 at 17.28truept
 \font\seventeeni=cmmi10 at 17.28truept
 \font\seventeenbf=cmbx10 at 17.28truept
 \font\seventeenit=cmti10 at 17.28truept
 \font\seventeensl=cmsl10 at 17.28truept
 \font\seventeensy=cmsy10 at 17.28truept
}
\def\loadfourteenpoint{
\font\fourteenrm=cmr10 at 14.4pt
\font\fourteeni=cmmi10 at 14.4pt
\font\fourteenit=cmti10 at 14.4pt
\font\fourteensl=cmsl10 at 14.4pt
\font\fourteensy=cmsy10 at 14.4pt
\font\fourteenbf=cmbx10 at 14.4pt
}
\def\loadtruetwelvepoint{
\font\twelverm=cmr10 at 12truept
\font\twelvei=cmmi10 at 12truept
\font\twelveit=cmti10 at 12truept
\font\twelvesl=cmsl10 at 12truept
\font\twelvesy=cmsy10 at 12truept
\font\twelvebf=cmbx10 at 12truept
}

\font\ninei=cmmi9
\font\eighti=cmmi8
\font\sixi=cmmi6
\skewchar\ninei='177 \skewchar\eighti='177 \skewchar\sixi='177

\font\ninesy=cmsy9
\font\eightsy=cmsy8
\font\sixsy=cmsy6
\skewchar\ninesy='60 \skewchar\eightsy='60 \skewchar\sixsy='60

\font\ninebf=cmbx9
\font\eightbf=cmbx8
\font\sixbf=cmbx6

\font\ninett=cmtt9
\font\eighttt=cmtt8

\hyphenchar\tentt=-1 % inhibit hyphenation in typewriter type
\hyphenchar\ninett=-1
\hyphenchar\eighttt=-1         

\font\ninesl=cmsl9
\font\eightsl=cmsl8

\font\nineit=cmti9
\font\eightit=cmti8

 % unslanted text italic
                      
\newskip\ttglue
\def\tenpoint{\def\rm{\fam0\tenrm}%
  \textfont0=\tenrm \scriptfont0=\sevenrm \scriptscriptfont0=\fiverm
  \textfont1=\teni \scriptfont1=\seveni \scriptscriptfont1=\fivei
  \textfont2=\tensy \scriptfont2=\sevensy \scriptscriptfont2=\fivesy
  \textfont3=\tenex \scriptfont3=\tenex \scriptscriptfont3=\tenex
  \def\it{\fam\itfam\tenit}\textfont\itfam=\tenit
  \def\sl{\fam\slfam\tensl}\textfont\slfam=\tensl
  \def\bf{\fam\bffam\tenbf}\textfont\bffam=\tenbf \scriptfont\bffam=\sevenbf
  \scriptscriptfont\bffam=\fivebf
  \normalbaselineskip=12pt
  \let\sc=\eightrm
  \let\big=\tenbig
  \setbox\strutbox=\hbox{\vrule height8.5pt depth3.5pt width\z@}%
  \normalbaselines\rm}

\def\twelvepoint{\def\rm{\fam0\twelverm}%
  \textfont0=\twelverm \scriptfont0=\ninerm \scriptscriptfont0=\sevenrm
  \textfont1=\twelvei \scriptfont1=\ninei \scriptscriptfont1=\seveni
  \textfont2=\twelvesy \scriptfont2=\ninesy \scriptscriptfont2=\sevensy
  \textfont3=\tenex \scriptfont3=\tenex \scriptscriptfont3=\tenex
  \def\it{\fam\itfam\twelveit}\textfont\itfam=\twelveit
  \def\sl{\fam\slfam\twelvesl}\textfont\slfam=\twelvesl
  \def\bf{\fam\bffam\twelvebf}\textfont\bffam=\twelvebf%
  \scriptfont\bffam=\ninebf
  \scriptscriptfont\bffam=\sevenbf
  \normalbaselineskip=12pt
  \let\sc=\eightrm
  \let\big=\tenbig
  \setbox\strutbox=\hbox{\vrule height8.5pt depth3.5pt width\z@}%
  \normalbaselines\rm}

\def\fourteenpoint{\def\rm{\fam0\fourteenrm}%
  \textfont0=\fourteenrm \scriptfont0=\tenrm \scriptscriptfont0=\sevenrm
  \textfont1=\fourteeni \scriptfont1=\teni \scriptscriptfont1=\seveni
  \textfont2=\fourteensy \scriptfont2=\tensy \scriptscriptfont2=\sevensy
  \textfont3=\tenex \scriptfont3=\tenex \scriptscriptfont3=\tenex
  \def\it{\fam\itfam\fourteenit}\textfont\itfam=\fourteenit
  \def\sl{\fam\slfam\fourteensl}\textfont\slfam=\fourteensl
  \def\bf{\fam\bffam\fourteenbf}\textfont\bffam=\fourteenbf%
  \scriptfont\bffam=\tenbf
  \scriptscriptfont\bffam=\sevenbf
  \normalbaselineskip=17pt
  \let\sc=\elevenrm
  \let\big=\tenbig                                          
  \setbox\strutbox=\hbox{\vrule height8.5pt depth3.5pt width\z@}%
  \normalbaselines\rm}

\def\seventeenpoint{\def\rm{\fam0\seventeenrm}%
  \textfont0=\seventeenrm \scriptfont0=\fourteenrm \scriptscriptfont0=\tenrm
  \textfont1=\seventeeni \scriptfont1=\fourteeni \scriptscriptfont1=\teni
  \textfont2=\seventeensy \scriptfont2=\fourteensy \scriptscriptfont2=\tensy
  \textfont3=\tenex \scriptfont3=\tenex \scriptscriptfont3=\tenex
  \def\it{\fam\itfam\seventeenit}\textfont\itfam=\seventeenit
  \def\sl{\fam\slfam\seventeensl}\textfont\slfam=\seventeensl
  \def\bf{\fam\bffam\seventeenbf}\textfont\bffam=\seventeenbf%
  \scriptfont\bffam=\fourteenbf
  \scriptscriptfont\bffam=\twelvebf
  \normalbaselineskip=21pt
  \let\sc=\fourteenrm
  \let\big=\tenbig                                          
  \setbox\strutbox=\hbox{\vrule height 12pt depth 6pt width\z@}%
  \normalbaselines\rm}

\def\ninepoint{\def\rm{\fam0\ninerm}%
  \textfont0=\ninerm \scriptfont0=\sixrm \scriptscriptfont0=\fiverm
  \textfont1=\ninei \scriptfont1=\sixi \scriptscriptfont1=\fivei
  \textfont2=\ninesy \scriptfont2=\sixsy \scriptscriptfont2=\fivesy
  \textfont3=\tenex \scriptfont3=\tenex \scriptscriptfont3=\tenex
  \def\it{\fam\itfam\nineit}\textfont\itfam=\nineit
  \def\sl{\fam\slfam\ninesl}\textfont\slfam=\ninesl
  \def\bf{\fam\bffam\ninebf}\textfont\bffam=\ninebf \scriptfont\bffam=\sixbf
  \scriptscriptfont\bffam=\fivebf
  \normalbaselineskip=11pt
  \let\sc=\sevenrm
  \let\big=\ninebig
  \setbox\strutbox=\hbox{\vrule height8pt depth3pt width\z@}%
  \normalbaselines\rm}

\def\eightpoint{\def\rm{\fam0\eightrm}%
  \textfont0=\eightrm \scriptfont0=\sixrm \scriptscriptfont0=\fiverm%
  \textfont1=\eighti \scriptfont1=\sixi \scriptscriptfont1=\fivei%
  \textfont2=\eightsy \scriptfont2=\sixsy \scriptscriptfont2=\fivesy%
  \textfont3=\tenex \scriptfont3=\tenex \scriptscriptfont3=\tenex%
  \def\it{\fam\itfam\eightit}\textfont\itfam=\eightit%
  \def\sl{\fam\slfam\eightsl}\textfont\slfam=\eightsl%
  \def\bf{\fam\bffam\eightbf}\textfont\bffam=\eightbf \scriptfont\bffam=\sixbf%
  \scriptscriptfont\bffam=\fivebf%
  \normalbaselineskip=9pt%
  \let\sc=\sixrm%
  \let\big=\eightbig%
  \setbox\strutbox=\hbox{\vrule height7pt depth2pt width\z@}%
  \normalbaselines\rm}

 % use after $ in ninepoint sections
\def\tenbig#1{{\hbox{$\left#1\vbox to8.5pt{}\right.\n@space$}}}
\def\ninebig#1{{\hbox{$\textfont0=\tenrm\textfont2=\tensy
  \left#1\vbox to7.25pt{}\right.\n@space$}}}
\def\eightbig#1{{\hbox{$\textfont0=\ninerm\textfont2=\ninesy
  \left#1\vbox to6.5pt{}\right.\n@space$}}}

% Page layout
%\newinsert\footins
\def\footnote#1{\edef\@sf{\spacefactor\the\spacefactor}#1\@sf
      \insert\footins\bgroup\eightpoint
      \interlinepenalty100 \let\par=\endgraf
        \leftskip=\z@skip \rightskip=\z@skip
        \splittopskip=10pt plus 1pt minus 1pt \floatingpenalty=20000
        \smallskip\item{#1}\bgroup\strut\aftergroup\@foot\let\next}
\skip\footins=12pt plus 2pt minus 4pt % space added when footnote is present
%\count\footins=1000 % footnote magnification factor (1 to 1)
\dimen\footins=30pc % maximum footnotes per page

\newinsert\margin
\dimen\margin=\maxdimen
%\count\margin=0 \skip\margin=0pt % marginal inserts take up no space

\loadtruetwelvepoint % At FNAL...
\loadtrueseventeenpoint

% \use\cs
% puts in the expansion of `\cs' if it's defined, the literal "\cs" otherwise.
\def\eatOne#1{}
\def\ifundef#1{\expandafter\ifx%
\csname\expandafter\eatOne\string#1\endcsname\relax}
\def\notTrue{\iffalse}\def\isTrue{\iftrue}
\def\ifdef#1{{\ifundef#1%
\aftergroup\notTrue\else\aftergroup\isTrue\fi}}
\def\use#1{\ifundef#1\linemessage{Warning: \string#1 is undefined.}%
{\tt \string#1}\else#1\fi}

%     \ref\label{text}
% generates a number, assigns it to \label, generates an entry.
% To list the refs on a separate page,  \listrefs
% \nref does the same without generating any text at the reference
% point
% June 26 1994: \preref postpones the generation of an entry, along with
% the text, until the first use of the reference

% 09/14/95: Added html...
%\def\hyperref#1#2{\special{html:<a href=\quote#1\quote>}{#2}\special{html:</a>}}
% 09/25/95: Now using Tanmoy Battacharya's macros...

%
\catcode`"=11
\let\quote="
\catcode`"=12
\chardef\foo="22
\global\newcount\refno \global\refno=1
\newwrite\rfile
\newlinechar=`\^^J
\def\@ref#1#2{\the\refno\n@ref#1{#2}}
% Added 09/14/95{\the\refno\n@ref#1{#2}}
\def\h@ref#1#2#3{\href{#3}{\the\refno}\n@ref#1{#2}}
\def\n@ref#1#2{\xdef#1{\the\refno}%
\ifnum\refno=1\immediate\openout\rfile=\jobname.refs\fi%
\immediate\write\rfile{\noexpand\item{[\noexpand#1]\ }#2.}%
\global\advance\refno by1}
\def\nref{\n@ref} % Hide to allow redefinitions of \ref,\nref to \preref
\def\ref{\@ref}   % without breaking the latter...
\def\hrref{\h@ref}
% To start a long reference...
\def\lref#1#2{\the\refno\xdef#1{\the\refno}%
\ifnum\refno=1\immediate\openout\rfile=\jobname.refs\fi%
\immediate\write\rfile{\noexpand\item{[\noexpand#1]\ }#2\semi}%
\global\advance\refno by1}
% To continue a long reference...
\def\cref#1{\immediate\write\rfile{#1\semi}}
% To end a long reference...

\def\preref#1#2{\gdef#1{\@ref#1{#2}}}

\def\semi{;\hfil\noexpand\break}

\def\listrefs{\vfill\eject\immediate\closeout\rfile%\parindent=20pt
\centerline{{\bf References}}\bigskip\frenchspacing%
\input \jobname.refs\vfill\eject\nonfrenchspacing}

\def\inputAuxIfPresent#1{\immediate\openin1=#1
\ifeof1\message{No file \auxfileName; I'll create one.
}\else\closein1\relax\input\auxfileName\fi%
}
% For references, some journal names

%and archives...

%\def\hepphref#1{\hyperref{http://xxx.lanl.gov/abs/hep-ph/#1}{archive}%
%{hep-ph/#1}{hep-ph/#1}}

%\def\hepphref#1{\href{http://xxx.lanl.gov/abs/hep-ph/#1}{hep-ph/#1}}

% An .aux file --- for forward references...
\newif\ifWritingAuxFile
\newwrite\auxfile
\def\SetUpAuxFile{%
\xdef\auxfileName{\jobname.aux}%
% Read it in if it exists
\inputAuxIfPresent{\auxfileName}%
% Now write a new one.
\WritingAuxFiletrue%
\immediate\openout\auxfile=\auxfileName}

% Some generally useful notation

% Warn about changed counters...

\catcode`\@=\active
\catcode`@=12  % No longer.
\catcode`\"=\active

\newread\epsffilein    % file to \read
\newif\ifepsffileok    % continue looking for the bounding box?
\newif\ifepsfbbfound   % success?
\newif\ifepsfverbose   % report what you're making?
\newdimen\epsfxsize    % horizontal size after scaling
\newdimen\epsfysize    % vertical size after scaling
\newdimen\epsftsize    % horizontal size before scaling
\newdimen\epsfrsize    % vertical size before scaling
\newdimen\epsftmp      % register for arithmetic manipulation
\newdimen\pspoints     % conversion factor
\pspoints=1bp          % Adobe points are `big'
\epsfxsize=0pt         % Default value, means `use natural size'
\epsfysize=0pt         % ditto
\def\epsfbox#1{\global\def\epsfllx{72}\global\def\epsflly{72}%
   \global\def\epsfurx{540}\global\def\epsfury{720}%
   \def\lbracket{[}\def\testit{#1}\ifx\testit\lbracket
   \let\next=\epsfgetlitbb\else\let\next=\epsfnormal\fi\next{#1}}%
\def\epsfgetlitbb#1#2 #3 #4 #5]#6{\epsfgrab #2 #3 #4 #5 .\\%
   \epsfsetgraph{#6}}%
\def\epsfnormal#1{\epsfgetbb{#1}\epsfsetgraph{#1}}%
\def\epsfgetbb#1{%
%
%   The first thing we need to do is to open the
%   PostScript file, if possible.
%
\openin\epsffilein=#1
\ifeof\epsffilein\errmessage{I couldn't open #1, will ignore it}\else
%
%   Okay, we got it. Now we'll scan lines until we find one that doesn't
%   start with %. We're looking for the bounding box comment.
%
   {\epsffileoktrue \chardef\other=12
    \def\do##1{\catcode`##1=\other}\dospecials \catcode`\ =10
    \loop
       \read\epsffilein to \epsffileline
       \ifeof\epsffilein\epsffileokfalse\else
%
%   We check to see if the first character is a % sign;
%   if not, we stop reading (unless the line was entirely blank);
%   if so, we look further and stop only if the line begins with
%   `%%BoundingBox:'.
%
          \expandafter\epsfaux\epsffileline:. \\%
       \fi
   \ifepsffileok\repeat
   \ifepsfbbfound\else
    \ifepsfverbose\message{No bounding box comment in #1; using defaults}\fi\fi
   }\closein\epsffilein\fi}%
%
%   Now we have to calculate the scale and offset values to use.
%   First we compute the natural sizes.
%
\def\epsfclipstring{}% do we clip or not?  If so,
\def\epsfsetgraph#1{%
   \epsfrsize=\epsfury\pspoints
   \advance\epsfrsize by-\epsflly\pspoints
   \epsftsize=\epsfurx\pspoints
   \advance\epsftsize by-\epsfllx\pspoints
%
%   If `epsfxsize' is 0, we default to the natural size of the picture.
%   Otherwise we scale the graph to be \epsfxsize wide.
%
   \epsfxsize\epsfsize\epsftsize\epsfrsize
   \ifnum\epsfxsize=0 \ifnum\epsfysize=0
      \epsfxsize=\epsftsize \epsfysize=\epsfrsize
      \epsfrsize=0pt
%
%   We have a sticky problem here:  TeX doesn't do floating point arithmetic!
%   Our goal is to compute y = rx/t. The following loop does this reasonably
%   fast, with an error of at most about 16 sp (about 1/4000 pt).
% 
     \else\epsftmp=\epsftsize \divide\epsftmp\epsfrsize
       \epsfxsize=\epsfysize \multiply\epsfxsize\epsftmp
       \multiply\epsftmp\epsfrsize \advance\epsftsize-\epsftmp
       \epsftmp=\epsfysize
       \loop \advance\epsftsize\epsftsize \divide\epsftmp 2
       \ifnum\epsftmp>0
          \ifnum\epsftsize<\epsfrsize\else
             \advance\epsftsize-\epsfrsize \advance\epsfxsize\epsftmp \fi
       \repeat
       \epsfrsize=0pt
     \fi
   \else \ifnum\epsfysize=0
     \epsftmp=\epsfrsize \divide\epsftmp\epsftsize
     \epsfysize=\epsfxsize \multiply\epsfysize\epsftmp   
     \multiply\epsftmp\epsftsize \advance\epsfrsize-\epsftmp
     \epsftmp=\epsfxsize
     \loop \advance\epsfrsize\epsfrsize \divide\epsftmp 2
     \ifnum\epsftmp>0
        \ifnum\epsfrsize<\epsftsize\else
           \advance\epsfrsize-\epsftsize \advance\epsfysize\epsftmp \fi
     \repeat
     \epsfrsize=0pt
    \else
     \epsfrsize=\epsfysize
    \fi
   \fi
%
%  Finally, we make the vbox and stick in a \special that dvips can parse.
%
   \ifepsfverbose\message{#1: width=\the\epsfxsize, height=\the\epsfysize}\fi
   \epsftmp=10\epsfxsize \divide\epsftmp\pspoints
   \vbox to\epsfysize{\vfil\hbox to\epsfxsize{%
      \ifnum\epsfrsize=0\relax
        \includegraphics{#1}%
      \else
        \epsfrsize=10\epsfysize \divide\epsfrsize\pspoints
        \includegraphics{#1}%
      \fi
      \hfil}}%
\global\epsfxsize=0pt\global\epsfysize=0pt}%
%
%   We still need to define the tricky \epsfaux macro. This requires
%   a couple of magic constants for comparison purposes.
%
{\catcode`\%=12 \global\let\epsfpercent=%\global\def\epsfbblit{%BoundingBox}}%
%
%   So we're ready to check for `%BoundingBox:' and to grab the
%   values if they are found.
%
\long\def\epsfaux#1#2:#3\\{\ifx#1\epsfpercent
   \def\testit{#2}\ifx\testit\epsfbblit
      \epsfgrab #3 . . . \\%
      \epsffileokfalse
      \global\epsfbbfoundtrue
   \fi\else\ifx#1\par\else\epsffileokfalse\fi\fi}%
%
%   Here we grab the values and stuff them in the appropriate definitions.
%
\def\epsfempty{}%
\def\epsfgrab #1 #2 #3 #4 #5\\{%
\global\def\epsfllx{#1}\ifx\epsfllx\epsfempty
      \epsfgrab #2 #3 #4 #5 .\\\else
   \global\def\epsflly{#2}%
   \global\def\epsfurx{#3}\global\def\epsfury{#4}\fi}%
%
%   We default the epsfsize macro.
%
\def\epsfsize#1#2{\epsfxsize}
%
%   Finally, another definition for compatibility with older macros.
%

%%%%%%%%%%%%%

%\magnification=\magstep1
\baselineskip=14pt

\rightline{ITP-SB-96-12}
\rightline{hep-th yymmdd}

\vglue .5cm
\centerline{\bf MULTI-MONOPOLE MODULI SPACES FOR} 
\centerline{\bf SU(N) GAUGE GROUP}

\vskip .3cm

\vglue 1cm
\centerline{{\bf Gordon Chalmers}}
\vskip .2 cm
\baselineskip=13pt
\centerline{\it Institute for Theoretical Physics} 
\centerline{\it State University of Stony Brook}
\centerline{\it Stony Brook, NY 11794-3840} 
\centerline{\it E-mail address : chalmers@insti.physics.sunysb.edu}

\vglue 2cm

\centerline{\bf ABSTRACT}
\vglue 0.3cm

The moduli space describing the low-energy dynamics of BPS multi-monopoles for several charge
configurations is presented. We first prove the conjectured form of the moduli space of $n-1$ distinct
monopoles in a spontaneously broken $SU(n)$ gauge theory. We further propose the solution where one of
the charge components has two units, hence asymptotically corresponds to embeddings of two monopoles
of one charge type and the rest different. The latter hyperk\"ahler metrics possess features of the
two-monopole Atiyah-Hitchin metric. We also conjecture classes of solutions to multi-monopole moduli
spaces with arbitrary charge and no more than two units in each component, which models the gluing
together of Atiyah-Hitchin metrics. Our approach here uses the generalized Legendre transform
technique to find the new hyperk\"ahler manifolds and rederive previously conjectured ones. 

\vfill
\eject

%%%%%%%%%%%%%%%%%%%%%%%%%%%%%%%%%%%%%%%%%%%%%%%%%%%%%%%%%%%%%%%%%%%% 

% \input texinclude/header.tex
% \input epsf
\SetUpAuxFile
\loadfourteenpoint
\hfuzz 60 pt
\FiguresInTexttrue
%\FiguresInTextfalse
%\draft

\def\c{\cdot}
\def\eps{\epsilon}

%%%%%%%%%%%%%%%%%%%%%%%%%%%%%%%%%%%%%%%%%%%%%%%%%%%%%%%%%%%%%%%%%%%%% 

% References

\preref\PapTown{ G.\ Papadopoulous, P.K.\ Townsend, Nucl.\ Phys.\ B444:245 (1995)}% 

\preref\Osborn{ H.\ Osborn, Phys.\ Lett.\ B83:321 (1979)}% 

\preref\WellSep{ G.W.\ Gibbons, N.S.\ Manton, hep-th/9506052}%

\preref\LongRange{ N.S.\ Manton, Phys.\ Lett.\ B154:5 (1985)}% 

\preref\PosAction{ G.W.\ Gibbons, C.N.\ Pope, Comm.\ Math.\ Phys.\ 66:267 (1979)}% 

\preref\AHSoln{ M.\ Atiyah, N.J.\ Hitchin, Phil.\ Trans.\ R.\ Soc.\ Lon.\ A315:459 (1985)}%

\preref\RemarkScatt{ N.S.\ Manton, Phys.\ Lett.\ B110:54 (1982)}% 

\preref\TopCharge{ E.\ Witten, D.\ Olive, Phys.\ Lett.\ B78:97 (1978)} 

\preref\DualPap{ C.\ Montonen, D.\ Olive, Phys.\ Lett.\ B72:117 (1977)} 

\preref\CHTone{ C.\ H.\ Taubes, Comm.\ Math.\ Phys.\ 80:343 (1981)} 

\preref\HkQC{ N.J.\ Hitchin, A.\ Karlhede, U.\ Lindstrom, M.\ Ro\v cek, 
 Comm.\ Math.\ Phys.\ 108:537 (1987)}

\preref\MantonMod{ N.S.\ Manton, Phys.\ Lett.\ B110:54 (1982)}% 

\preref\AHBook{ M.\ Atiyah, N.J.\ Hitchin, The geometry and dynamics of magnetic 
monopoles, Princeton University Press, NJ (1988)} 

\preref\TwistLT{ Ivan.\ T.\ Ivanov, M.\ Ro\v cek, hep-th/9512075} 

\preref\GLT{ U.\ Lindstrom, M.\ Ro\v cek, Comm.\ Math.\ Phys.\ 115:21 (1988)}

\preref\GNO{ P.\ Goddard, J.\ Nuyts, D.\ Olive, Nucl.\ Phys.\ B125:1 (1977)}% 

\preref\Connell{ S.A.\ Connell, unpublished}%

\preref\Sen{ A.\ Sen, Phys.\ Lett.\ B329:217 (1994)}% 

\preref\WeinWellSep{ K.\ Lee, E.J.\ Weinberg, P.\ Yi, hep-th/9602167} 

\preref\WeinBigGroup{ K.\ Lee, E.J.\ Weinberg, P.\ Yi, hep-th/9601097} 

\preref\WeinCountone{ E.J.\ Weinberg, Phys.\ Rev.\ D20:936 (1979)} 

\preref\WeinFundArbGroup{ E.J.\ Weinberg, Nucl.\ Phys.\ B167:500 (1980)} 

\preref\Harvey{ Jeffrey\ A.\ Harvey, hep-th/9603086}

\preref\GauntlettBigGroup{ J.P.\ Gauntlett, D.A.\ Lowe, hep-th/9601085} 

\preref\CollQuant{ J.P.\ Gauntlett, Nucl.\ Phys.\ B411:443 (1994)} 

\preref\Blum{ J.\ Blum, Phys.\ Lett. B333:92 (1994)}

\preref\TwoMonople{ G.W.\ Gibbons, N.S.\ Manton, Nucl.\ Phys.\ B274:183 (1986)} 

\preref\BPSone{ E.B.\ Bogomol'nyi, Sov.\ J.\ Nucl.\ Phys.\ 24:449 (1976)} 

\preref\BPStwo{ M.K.\ Prasad, C.H.\ Sommerfield, Phys.\ Rev.\ Lett.\ 35:760 (1975)} 

\preref\Callias{ C.\ Callias, Comm.\ Math.\ Phys.\ 62:213 (1978)} 

\preref\progress{ G.\ Chalmers, in progress}%

\preref\projections{ M.K.\ Murray, Comm.\ Math.\ Phys. 96:539 (1984)}% 

%%%%%%%%%%%%%%%%%%%%%%%%%%%%%%%%%%%%%%%%%%%%%%%%%%%%%%%%%%%%%%%%%%%%%%%%%% 

\noindent{\bf I. Introduction}
\vskip .3in

Recently there has been much interest in the study of monopole interactions in the BPS limit. For some
time it has been known that the Yang-Mills-Higgs field equations admit solitonic solutions
[\use\BPSone,\use\BPStwo], and that there are in general a finite dimensional solution space labelled
by the same magnetic charge [\use\WeinCountone,\use\CHTone]. These solutions correspond in the
asymptotic regions to the superposition of $n$ fundamental monopoles. The non-trivial global geometry
of these moduli spaces is of interest for several reasons. The explicit form of the metric allows one
to determine the low-energy dynamics of interacting monopoles [\use\MantonMod]. Furthermore, a
collective coordinate quantization of these low-energy degrees of freedom allow one to investigate the
existence of stable bound state solutions, which gives a non-trivial test of the existence of
non-perturbative states predicted by the conjectured
$SL(2,Z)$ duality of $N=4$ supersymmetric Yang-Mills theory.

The $SU(2)\rightarrow U(1)$ charge $2$ monopole problem has been extensively studied in the past and
it is known that in general the moduli spaces have a hyperk\"ahler structure
[\use\AHBook,\use\AHSoln]. Even for larger broken gauge groups, the different charge two centered
moduli spaces are unique in that the possible candidate four-dimensional hyperk\"ahler manifolds,
those containing a rotational $SU(2)$ or $SO(3)$ group of isometries and/or additional charge
conservation factor $U(1)$, have been completely classified [\use\AHBook]. By imposing the appropriate
physical isometries one may directly find the correct moduli space. No such classification is known of
the hyperk\"ahler manifolds which describe charge $k>2$ moduli spaces, although the asymptotic form of
the metric is known [\use\WellSep]. 

Monopoles in gauge groups such as $SU(n)$ have charges associated with each of the Cartan generators
(or rather labelled according to the dual of the root lattice), and for widely separated
configurations the gauge fields are built up from embedding independent $SU(2)$ fundamental monopoles
[\use\WeinFundArbGroup]. There are $n-1$ independent fundamental charge one monopole solutions arising
within the BPS equations for the case of a maximally broken $SU(n)$ theory. The moduli space for $k<n$
distinct monopole gauge configurations contains an additional set of commuting $U(1)^k$ isometries and
has been analytically found in the asymptotic region where the monopoles are all widely separated from
one another [\use\WeinWellSep]. The asymptotic form of the metric does not develop singularities as
one continues into the core region and has been conjectured to be correct throughout
[\use\WeinWellSep]. Further examples of all-$k$ charge moduli spaces corresponding to mixed monopole
configurations in $SU(n)$ have not appeared in the literature and will be presented in this paper. 

The Legendre transform construction of hyperk\"ahler manifolds and its extensions is naturally suited
to the construction of multi-monopole moduli spaces [\use\HkQC,\use\GLT]. This formalism has recently
been used, for example, to construct Atiyah-Hitchin metric ${\tilde M}_2$, which describes the moduli
space of a charge two monopole in a broken $SU(2)$ theory [\use\TwistLT]. These techniques will be
used extensively in the construction of moduli spaces presented this paper. The most attractive
feature of this type of construction when applied to the monopole moduli space problem is that it
allows one to generalize in a straightforward manner entire classes of moduli spaces; These examples
are conjectured to give the moduli spaces of monopoles in higher gauge groups with no more than two
units of charge along a particular Cartan direction, corresponding to gluing together many
Atiyah-Hitchin metrics.

In this paper we propose the form of the moduli space for various multi-monopole solutions to the BPS
equations in an $SU(n)$ gauge theory broken down to $U(1)^{n-1}$. In section II we review some
defining properties of BPS monopoles and the asymptotic form of the general multi-monopole moduli
space. In section III we present the Legendre transform technique and illustrate its use in working
with $SU(n)$ moduli spaces by constructing the moduli space of up to $n-1$ distinct monopoles. The
metric on the space corresponding to this charge configuration has been recently conjectured to be the
same as its asymptotic form, and we prove this to be correct. In Section IV we present the form for
the moduli space ${\tilde M}_k$ for any number of monopoles where two are like charged and the rest
different. In Section V we conjecture further examples of moduli spaces where no more than two
monopoles are labelled by the same magnetic quantum number. In Section VI we conclude with some
details about related problems. 

\vskip .4in
\noindent{\bf II. BPS Monopoles}
\vskip .3in

\noindent{\it BPS Equations and Fundamental Monopoles} \vskip .2in

In this section we review the properties of the BPS equations in an $SU(n)$ gauge theory. We define as
usual the raising and lowering operators of $su(n)$ as

$$
\left[ H_i , H_j \right] = 0 \quad\quad
\left[ E_{\vec{\alpha}}, E_{-\vec{\alpha}} \right] = \vec{\alpha} \c \vec{H} $$
$$
\left[ E_{\vec{\alpha}}, H \right] = \vec{\alpha} E_{\vec{\alpha}} \ . \anoneqn
$$ where $\vec{\alpha}$ are the roots. The Cartan generators are normalized so that ${\rm Tr}H_i H_j =
\delta_{ij}$. Any vector $\vec{\alpha}$ on the root lattice can be decomposed into a basis of simple
roots
$\vec{\alpha}_i$ with integral coefficients

$$
\vec{\alpha} = \sum_{i=1}^r n_i \vec{\alpha}_i , \quad n_i \geq 0 \ . \anoneqn
$$ One may define the dual lattice spanned by the vectors
$\vec{\alpha}^\ast_i = {\vec{\alpha}_i\over \alpha_i^2}$, 

$$
\vec{\mu} = \sum_{j=1}^r n_i^\ast \vec{\alpha}_j^\ast \ . \anoneqn
$$ For $SU(n)$ the simple roots obey $\vec{\alpha}_i \c \vec{\alpha}_i >0$ and $\vec{\alpha}_i \c
\vec{\alpha}_j <0$ for $i\neq j$. The signs will turn out to be very important in the construction of
the moduli space metrics.

\iffalse

In a maximally broken theory $SU(n)\rightarrow U(1)^{n-1}$ the states are labelled by the quantum
numbers $(n_i, n_j^\ast)$. The topological quantization condition demands that the magnetic and
electric charges in the basis above are labelled by the roots and their duals $\vec{q}_e =
e\vec{\alpha}$ and $\vec{q}_m = {4\pi\over e}
\vec{\alpha}^\ast$ [\use\GNO]. The central charges are defined by the surface integrals at infinity, 

$$ Q_e = {1\over v} \int_{S^2} d{\hat\Sigma} \c {\rm Tr}(\vec{E} \Phi) \quad\quad Q_m = {1\over v}
\int_{S^2} d{\hat\Sigma} \c {\rm Tr}(\vec{B} \Phi) \ , \anoneqn
$$ where $v$ is the magnitude of the vev of the Higgs field at infinity, ${\rm Tr}\Phi\Phi = v^2$.

Classically the spectrum of fundamental particles (i.e. $W^{\pm}$ analogs) consist of the electrically
charged states with no magnetic charge and electric charges $n_i = \pm,0$. The conjectured $SL(2,Z)$
duality predicts towers of dual stable magnetically charged states in an
$N=4$ theory with non-zero $n_j^\ast$ [\use\DualPap,
\use\Osborn]; these states are believed to be found semi-classically by investigating the moduli space
of magnetically charged solutions to the BPS equations [\use\Sen]. 

The charge vectors obey the topological quantization condition $\vec{q}_e \c \vec{q}_m = 4\pi n$, with
$n$ an integer. The tower of BPS saturated states satisfy the classical mass formula $m =
\vert \vec{h}\c\vec{q}_e + (\vec{h}\c \vec{q}_m) \vert$, where $h$ is a root vector defined by the
Higgs field evaluated asymptotically in some direction as $\Phi = \vec{h} \c \vec{H}$. 

\fi

Monopoles in the BPS limit are soliton solutions to the Yang-Mills-Higgs field equations in which the
presence of the scalar potential is replaced with boundary conditions on the fields at spatial
infinity. We have the Lagrangian

$$ {\cal L} = -{1\over 4} {\rm Tr} F_{\mu\nu} F^{\mu\nu} + {1\over 2}  {\rm Tr} D_\mu \Phi D^\mu
\Phi - V(\Phi) \ ,
\eqn\monaction
$$ with

$$
\eqalign{ & F_{\mu\nu} = \partial_\mu A_\nu - \partial_\nu A_\mu + e \left[ A_\mu ,  A_\nu \right]
\cr & D_\mu \Phi = \partial_\mu \Phi + e \left[ A_\nu , \Phi \right] }
$$ and scalar potential

$$ V(\Phi) = \lambda {\rm Tr} (\Phi^2 - v^2)^2 \anoneqn
$$ All fields are taken in the adjoint representation of $SU(n)$. Finite energy configurations
necessarily satisfy the boundary condition that at spatial infinity $\Phi^2 \rightarrow v^2$. 

The equations of motion from the Lagrangian in eqn.(\use\monaction) are

$$ D_\mu F^{\mu\nu} = e \left[ \Phi, D^\nu \Phi \right] \quad\quad D^\mu D_\mu \Phi = -4 \lambda
\Phi(\Phi^2 -v^2)
\, \anoneqn
$$ together with the Bianchi identity. The first equation for the case of a configuration in the gauge
$A_0 = 0$ gives Gauss' law, 

$$ D_i F^{i0} = e \left[ \Phi, D^0 \Phi \right] \ . \anoneqn
$$ In the BPS limit we take
$\lambda \rightarrow 0$, while maintaining the asymptotic condition on the scalar field expectation
value. The potential is reflected then as a topological boundary condition on the scalar and gauge
fields on the sphere at spatial infinity $S^2$. It is well known that the topological sectors of the
gauge field configurations for a simply connected group spontaneously broken from $G$ to $H$ are
labelled by $\pi_2 (G/H)$, which for $SU(n)
\rightarrow U(1)^{n-1}$ is $\pi_1 (H) = Z^{n-1}$ (has dimension $n-1$). 

The static multi-monopole gauge solutions are found by first looking for the minimal energy
configurations corresponding to the Lagrangian in eqn.(\use\monaction). From the colored electric and
magnetic fields $E^a_i = F^a_{0i}$ and $B_i^a = {1\over 2} \eps_{ijk} F^{jk,a}$, the static energy is 

$$ {\cal H} = {1\over 4} {\rm Tr}~\int d^3x~ (B_i + D_i\Phi)^2 +  (B_i - D_i\Phi)^2 + 4 V(\Phi) \ .
\anoneqn
$$ In the limit in which the scalar self-coupling $\lambda$ is taken to zero, the minimal energy
configurations to the BPS equations [\use\BPSone,\use\BPStwo] satisfy 

$$ D_i \Phi = \pm {1\over 2} \eps_{ijk} F^{jk} \ , \anoneqn
$$ in addition to a boundary condition on the fields at infinity which represent the non-trivial
homotopy class of the gauge field. We demand the field configurations to have finite energy. They must
satisfy the asymptotic conditions 

$$
\eqalign{ &
\Phi = \vec{h} \c \vec{H} + {1 \over 4\pi r} \vec{q}_m \c \vec{H}  + {\cal O}({1\over r^2})
\cr &
\vec{B} = {\hat{r}\over 4\pi r^2} \vec{q}_m \c\vec{H}  + {\cal O}({1\over r^3}) \ , }
\eqn\topcond
$$ where the Higgs field vacuum expectation value defined at infinity along some direction is given by
$\lim_{\vert\vec{r}\vert\rightarrow\infty} \langle\Phi\rangle\ = \vec{h} \c \vec{H}$. In order for the
gauge group to be completely broken we require $\vec{h} \c \vec{\alpha}_i \neq 0$. There exists then a
basis of the simple roots such that $\vec{h} \c \vec{\alpha}_i >0$, which is what we take in the
remainder of this paper [\use\WeinFundArbGroup].

In a maximally broken theory $SU(n)\rightarrow U(1)^{n-1}$ the states are labelled by the quantum
numbers $(n_i, n_i^\ast)$. The topological quantization condition demands that the magnetic and
electric charges in the basis above are labelled by the root lattice and its dual $\vec{q}_e =
e\vec{\alpha}$ and $\vec{q}_m = {4\pi\over e}
\vec{\alpha}^\ast$ [\use\GNO]. The charge vectors then obey the condition $\vec{q}_e \c \vec{q}_m =
4\pi n$, with
$n$ an integer. By expanding the fields about the broken vacuum we find the spectrum of elementary
particles (i.e.
$W^{\pm}$ analogs) consist of the electrically charged states with no magnetic charge and electric
charges in one to one correspondence with the roots. (By convention we will use $k^i$ rather than
$n_i^\ast$, the set of which we label by $\vec{k}$.) 

The BPS masses of the monopole states are given by $m = \vert \vec{h} \c \vec{q}_m \vert$ with
$\vec{q}_m = {4\pi
\over e} \sum_j^{n-1} k^i \vec{\alpha}_i^\ast$. An important difference for multi-monopoles in the
larger gauge groups is the existence of neutrally stable states. In the case of $SU(n)$ with
$n>2$, the masses of states based on the composite roots $\vec{\alpha}^\ast = \sum_j k^i
\vec{\alpha}_i^\ast$ saturate the triangle equality; the mass of the monopole decomposes as a sum of
the fundamental masses $m= \sum_j k_j m_j$ with $m_j = {4\pi\over e} \vec{h} \c
\vec{\alpha}_j^\ast$ (recall $\vec{h}
\c \vec{\alpha}_j >0$). The monopole based on non-simple roots is then neutrally stable to decay into
single ones based on the simple roots. 

Considering the Lagrangian in eqn.(\use\monaction) as the bosonic part of an $N=4$ supersymmetric
model, which then requires additional adjoint scalars $\Phi^I$ together with replacing the potential
by ${\rm Tr}[\Phi^I,\Phi^J]^2$, the conjectured $SL(2,Z)$ duality predicts towers of dual, stable
magnetically charged states with non-zero $k_j$ [\use\DualPap, \use\Osborn] satisfying the BPS mass
formula. Since the mass formula is exact in this case, these states are believed to be found
semi-classically by investigating the moduli space of magnetically charged solutions to the BPS
equations [\use\Sen]. This provides a strong motivation for studying the moduli space of magnetically
charged states.

Before discussing the moduli space in more detail, it is important to understand the asymptotic regime
where monopoles are widely separated from one another. This region gives an intuitive understanding of
the parameter counting for the dimension of the various moduli spaces, and furthermore in this regime
the forces between monopoles can be found to leading order.

E. Weinberg has shown how multi-monopole solutions for higher gauge groups completely broken down to
its maximal torus may be understood by embedding independent $SU(2)$ solutions along the simple root
directions in $SU(n)$ [\use\WeinFundArbGroup]. We first pick a root vector $\vec{\alpha}$ in
$SU(n)$; along this direction an explicit
$SU(2)$ sub-algebra is generated by

$$ T_1 = {1\over \sqrt{2 \vec{\alpha}^2}} (E_{\vec{\alpha}} + E_{-\vec{\alpha}}) \quad\quad T_2 =
{-i\over \sqrt{2
\vec{\alpha}^2}} (E_{\vec{\alpha}} - E_{-\vec{\alpha}}) \quad\quad T_3 = {1\over \vec{\alpha}^2}
\vec{\alpha} \c
\vec{H} \ . \eqn\subgen
$$ Let $\Phi^j_{\rm{su}(2)}$ and $A^j_{\rm{su}(2)}$ be a charge $k$, $SU(2)$ solution ($j$ labels the
generators in eq.(\use\subgen) and $\vec{\Phi}_{\rm{su}(2)} \c \vec{T}\equiv \sum_{j=1}^3
\Phi^j_{\rm{su}(2)} T_j$) with the vacuum expectation at infinity for the Higgs given by
$v_{\rm{su}(2)} = \vec{h} \c \vec{\alpha}$. 

An explicit solution for embedding the $SU(2)$ multi-monopole is given by 

$$
\eqalign{ &
\Phi = \vec{\Phi}_{\rm{su}(2)} \c \vec{T} + \vec{H} \c 
\Bigl( \vec{h} - \vec{\alpha} ~{\vec{h} \c \vec{\alpha}^\ast} \Bigr) \cr & A =
\vec{A}_{\rm{su}(2)} \c \vec{T}
\cr & v = \vec{h} \c \vec{\alpha} \ . }
\eqn\embedsoln
$$ The magnetic charge of the BPS solution (\use\embedsoln) is given along the dual to the embedding
root and is
$\vec{q}_m = {4\pi\over e} \vec{\alpha}^\ast$ (with mass $m={4\pi\over e} \vert \vec{h}\c
\vec{\alpha}^\ast \vert$) . One has picked a particular $U(1)$ factor, i.e. one vector
$\vec{\alpha}$, in providing this solution.

With the embedding solutions (\use\embedsoln) and the basis of simple roots one may introduce the
``fundamental'' charge $\vert \vec{k}\vert =1$ monopoles, each of which live along different simple
root directions. In general a multi-monopole solution can not be thought of as a sum of individual BPS
solutions; however, in the asymptotic regime where we keep only the leading $1/r$ terms in the gauge
fields, the identification can be made precise. To every simple root vector
$\vec{\alpha}_i$ we associate an $SU(2)$ embedding via eqn.(\use\embedsoln) with only one unit of
magnetic charge. For widely separated configurations the general multi-monopole may be thought of as a
linear superposition of the fundamental ones, each with charge one and possibly along different weight
directions. 

Asymptotically any $SU(n)$ monopole is a linear superposition of the fundamental monopoles (of
individual charge one) and the centered moduli space approaches a product manifold
${\tilde {\cal M}}_{\vec{k}} \rightarrow \bigl({\tilde{\cal M}}_1 \bigr)^{\sum_i k_i}$. Furthermore,
from the number of zero modes of a single $SU(2)$ monopole, $3$ $\delta \vec{r}_i$ translational and
one $U(1)$ rotational, one may roughly obtain the dimension of the moduli space for the general
multi-monopole solutions. Given that the moduli space is smooth and simply connected, $d$ does not
change; naively we obtain
$N=\sum_{i=1}^{n-1} 4 k^i$ as the number of zero modes for the general monopole. Although this is a
rather intuitive argument, this number of moduli parameters may be derived rigorously
[\use\WeinCountone,
\use\WeinFundArbGroup]. 

\vskip .3in
\noindent{\it Moduli Spaces}

In general there is a multi-parameter family of solutions to the BPS equations with the same
topological charge. The moduli space of a monopole configuration is found by considering the
deformations of the fields in the non-trivial gauge background which are not gauge equivalent and
preserve the equations of motion. By definition
${\cal M}_{\vec{k}}$ is defined as the space ${\cal G}$ of all solutions to the BPS equations of
constant potential energy modulo the gauge deformations ${\cal H}$ connected to the identity. In this
section we review some properties of these spaces. 

The zero modes of the multi-monopole solution are found by considering a small field deformation of
the BPS equations around a charge $\vec{k}$ solution. As usual, the $A_0 =0$ gauge condition is taken
to fix the residual background gauge invariances connected to the identity. To linear order, the
perturbations of the fields satisfy 

$$
\eqalign{ & A^k = {\tilde A}^k + \delta A^k(\vec{x},t) \quad\quad \Phi = {\tilde\Phi} + 
\delta\Phi (\vec{x},t)
\cr &
\eps_{ijk} {\tilde D}_j (\delta A_k) = {\tilde D}_i (\delta \Phi)  + \left[ \delta A_i, {\tilde
\Phi} \right] \ , }
\eqn\BPSdeform
$$ where Gauss's law comes from the equations of motion, 

$$ {\tilde D}_i (\delta \partial_0 A^i) = e \left[ {\tilde\Phi}, 
\delta\partial_0 \Phi \right] \ .
\anoneqn
$$ The latter equation insures that the kinetic energy remains zero. There are remaining $U(1)$ gauge
invariances acting on the perturbations from rotating $\delta A_k$ which we eliminate (since the
moduli space is defined as the space of solutions modulo small gauge transformations) by imposing the
gauge fixing on the fluctuating fields

$$ {\tilde D}_i (\delta A^i) + \left[ {\tilde\Phi}, \delta \Phi \right] =0 \eqn\resfix
$$ In order to arrive at the low-energy dynamics we allow the small deformations $\delta A_k$ and
$\delta\Phi$ to vary with time and then search for the most general solution to the above equations. 

Consider for example a charge $\vert\vec{k}\vert=1$ monopole solution, then constant spatial
translations of the
$SU(2)$ embeddings in eq.(\use\embedsoln) do not change the BPS equations and generate a set of three
zero-modes, one for each direction in $R^3$. Time dependent spatial translations however raise the
kinetic energy. In the case of higher charge solutions, where we do not have a notion of separating a
monopole into charge one components except in the asymptotic regions, this kinetic energy contribution
is found by quantizing the moduli of the BPS equations.

There are, however, further low-energy collective coordinates. Consider again the case of a
fundamental charge
$\vert\vec{k}\vert =1$ monopole. The linearized BPS equations and the gauge fixing term in
eqn.(\use\resfix) only freeze out small gauge transformations. The remaining collective coordinate is
found by looking at pure gauge deformations of the fields which in general give rise to a time
dependence in the small perturbations. In the
$A_0=0$ background gauge we have, 

$$
\partial_0 A_i = \delta A_i = D_i \Lambda \quad\quad \partial_0 \Phi = \delta\Phi = \left[ \Phi,
\Lambda \right]
\eqn\timegauge
$$ In this case the time dependence is pure gauge, and the potential remains constant in time - so
that any solution of the form (\use\timegauge) is a zero mode. The unique solutions to
eqns.(\use\timegauge , \use\resfix ,
\use\BPSdeform) for time dependent $\Lambda$ is given by $\Lambda = \chi (t) \Phi(\vec{r})$
parameterized by the large gauge transformation $g = {\rm e}^{\chi (t) \Phi(\vec{r})}$. In general the
parameter $\chi$ is allowed a smooth time dependence ($\chi$ is not allowed in general to vanish at
infinity). Note that $g$ is spatially dependent through the dependence in the scalar field, but does
not vanish at $\vert \vec{r}\vert \rightarrow\infty$ because of the boundary condition on the vev of
the scalar.

If $\partial_0 \chi=0$ then this transformation parameterizes a large gauge transformation, and
$\chi$ is a physical zero mode. The field dependence is $\partial_0 A_i = \delta A_i = D_i (\chi
\Phi), \delta\Phi =0, \delta A_0 =0$. The large global gauge transformations parameterized by
$\chi$ lives in the unbroken $U(1)$, and as a result, $\chi$ is a periodic coordinate ($\chi\in S^1$).
Time dependent solutions to $\chi$ increase the kinetic energy (keeping the potential energy fixed)
and gives the monopole fields electric charge. 

For the single monopole we have a total of four moduli, the translations and the electric charge of
the monopole, and the moduli space is $M_1 = R^3\otimes S^1$. Higher charge monopole configurations
give a different moduli space structure in $SU(n)$ for $n>2$. In $SU(2)$, the overall charge
conservation, generated by the global rotation above, together with the center of mass coordinates
factored out give a $4k$ dimensional space of the form [\use\AHBook], 

$$ M_k = R^3 \otimes {S^1\otimes {\tilde {\cal M}}_k \over Z_k} \ . \anoneqn
$$ The total moduli space is obtained by identifying the periodic values of the quantized charge
degrees of freedom, and $Z_k$ is a discrete subgroup of the isometry group of the centered moduli
space which enforces this identification.

In higher gauge groups the moduli space does not necessarily factorize onto an $S^1$ since the total
charge does not have to be periodic [\use\WeinBigGroup, \use\GauntlettBigGroup]. Consider
$SU(n)$ and let $q_i$ denote the quantized electric charges along the different $U(1)$ factors. The
center-of-mass coordinate corresponding to the total charge has the form 

$$ q = {1\over m e} \sum_{i=1}^{n-1} m_i q_i \quad\quad  m=\sum_{i=1}^{n-1} m_i \ ,
\anoneqn
$$ where $m_i = {4\pi\over e} k_i \vec{h} \c \vec{\alpha}_i$, and is not quantized unless all ratios
$m_i /m_j$ are rational. In this case the moduli space has the form, 

$$ M_{\vec{k}} = R^3 \otimes {R^1 \otimes {\tilde{\cal M}}_{\{\vec{k}\}} \over {\cal D}} \ ,
\anoneqn
$$ where $D$ is again a discrete subgroup of the isometry group of the centered moduli space. In this
work we only consider the construction of the (covered) centered moduli spaces ${\tilde{\cal
M}}_{\vec{k}}$ which asymptotically resemble a product of charge one $M_1$ factors.

We now briefly discuss the construction of the moduli spaces. It is well-known that the static BPS
equations are equivalent to the self-dual Yang-Mills equations in Euclidean space. Define ${\hat A}_i
= A_i$ and ${\hat A}_4 =
\Phi$. Equations (\use\BPSdeform) and (\use\resfix) may be combined into the form 

$$ G_{ij} = F_{ij} \quad\quad G_{i4} = D_i \Phi \, $$
$$ D_{[i} (\delta {\hat A}_{j]}) = \eps_{ij}^{~~kl} D_{[k}  (\delta {\hat A}_{l]}) \,
\anoneqn
$$
$$ D_i (\partial_0 A^i) = 0 \ ,
\anoneqn
$$ so that $G_{ij}={1\over 2} \eps_{ijkl} G^{kl}$. Monopoles then correspond to self-dual gauge
potentials satisfing the asymptotic conditions in eqn.(\use\topcond).

The slow time dependence is introduced by letting the gauge fields depend on the $N_{\{ \vec{k}\} }$
time dependent moduli parameters $y_i (t)$, $A_\mu \equiv A_\mu (\vec{r}_i , y_i(t))$. By definition,
the potential energy is independent of the moduli. Under a small perturbation of the gauge field in
the moduli direction $\kappa_i$, the fields change as,

$$ A_\mu \rightarrow A_\mu + \kappa^i (\delta_i A_\mu) \ . \anoneqn
$$ However, the perturbations $\delta_i A_\mu$ need to be constrained first by the background gauge
fixing before we obtain the tangent vectors to ${\cal M}_{\vec{k}}$ since gauge transformations of
$\delta_i A_\mu$ have been modded out. With the gauge fixing term $D_i (\delta A^i)=0$, the tangent
vectors are obtained by differentiating with respect to the moduli and compensating (if necessary) by
a small gauge transformation to keep the gauge fixing term intact. We have

$$
\delta_i A_\mu = {\partial A_\mu\over \partial y^i(t)} + (D_\mu {\tilde\Lambda}^i)
\anoneqn
$$ where $\tilde\Lambda^i (\vec{r}_j, y_j)$ is chosen to maintain the gauge-fixing in
eqn.(\use\resfix). This definition of tangent vectors appropriately corresponds to the moduli space,
which is defined by the charge
$\vec{k}$ solution set of the BPS equations modulo small gauge transformations. 

The inner product of the tangent vectors $\delta_i A_\mu^a$ defines the metric on the moduli
space $M_{\vec{k}}$ to be

$$ 
{\cal G}_{ij} = - \int d^3x {\rm Tr} \bigl( 
\delta_i A_\mu \delta_j A^\mu \bigr) \ . 
\anoneqn
$$ In order to arrive at the low-energy effective Lagrangian we make explicit the time dependence of
the moduli parameters (in $A_0$ gauge). 

$$
\partial_0 A_i = (\partial_0 y^i) \delta_i A_\mu \quad\quad \partial_i A_0 =0 \anoneqn
$$ and the time dependent action gives,

$$ S= -{1\over 2} \int d^4r {\rm Tr} F_{\mu\nu} F^{\mu\nu}  = {1\over 2} \int dt G_{ij} \partial_0 y^i
\partial_0 y^j \anoneqn
$$ The static portion of the action drops out since the fields satisfy the BPS equations.

The inter-monopole forces have recently been deduced in the asymptotic regime for any number of
separated monopoles in the higher rank gauge groups [\use\WeinBigGroup]. The forces give the moduli
space geometry of a higher-dimensional $d=4k$ analog of the Taub-NUT space, and shows the asymptotic
moduli space for $n$ fundamental monopoles possesses a metric which does not lead to singularities as
one brings them close together. This behavior is in contrast to the multi-monopole moduli space in
which the some of the asymptotic monopole fields are labelled along the same direction; in this case
the asymptotic form develops singularities, which are ultimately resolved by a charge exchange
process, as one continues it into the core region.

The space ${\cal M}_{\vec{k}}$ has been studied extensively for the gauge group $SU(2)$, and in
general is hyperk\"ahler and complete for all gauge groups. A natural language for the construction of
hyperk\"ahler manifolds is the Legendre transform (LT) and its generalizations [\use\TwistLT]. In the
following we will construct the ``distinct'' moduli spaces in the LT and generalize to further cases. 

\vskip .4in
\noindent{\bf III. Generalized Legendre Transform} \vskip .3in

The Legendre transform (LT) technique has been a powerful construction to locally generate
hyperk\"ahler manifolds from holomorphic functions [\use\HkQC,\use\GLT]. Although this construction
gives in general incomplete manifolds, it can in principle be used to generate the globally
well-defined spaces which are the hyperk\"ahler moduli spaces of the BPS equations. In the following
we review the LT formalism, and reformulate the moduli space Kahler potentials for the distinct
fundamental multi-monopole space.

Hyperk\"ahler manifolds are Riemannian manifolds which are K\"ahler with respect to three covariant
complex structures, denoted conventionally by $I$, $J$, and $K$ ($\nabla I=0$ et. al. with respect to
the Levi-Cevita connection). They obey the quaternionic algebra 

$$ I^2 = J^2 = K^2 = -1 \quad\quad IJ = K = -JI \ , \anoneqn
$$ and cyclic permutations. The three Kahler forms are then given by $\omega_I (X,Y)= g(IX,Y)$, etc.
An isometry is called tri-holomorphic if it preserves all three K\"ahler forms. 

We first define the holomorphic polynomials of order $2n$ over ${\rm CP}_1$ as 

$$
\eta^{(2n)}(\rho) = \sum_{j=0}^{2n} w_j \rho^j \ . \eqn\etafunctions
$$ The functions $\eta^{(2n)}(\rho)$ are sections of line bundles ${\cal O}(2n)$ over ${\rm CP}_1$
consisting of all polynomials of order $2n$, and $\rho$ is the ${\rm CP}_1$ coordinate. 

The standard representation of ${\rm CP}_1$ is found by pasting together two copies of the complex
plane $\Phi$,
$\tilde\Phi$ with coordinates $\rho$ and $\tilde\rho$; on the overlap $\Phi\cap{\tilde\Phi}$ the
coordinates are related by $\rho=1/{\tilde\rho}$. Furthermore, on the overlap of the two charts the
functions in eqn.(\use\etafunctions) transform as ${\tilde \eta}^{(2n)}(1/\rho) = \rho^{-2n}
\eta^{(2n)}(\rho)$. 

We impose a reality constraint on $\eta^{(2n)}$ given by 

$$ {\bar \eta^{(2n)}(\rho)} = (-1)^n {\bar\rho}^{2n} \eta^{(2n)}  (-1/{\bar\rho}) \ ,
\eqn\reality
$$ which means that $w_j = (-)^{n+j} {\bar w}_{2n-j}$. The operation in eqn.(\use\reality) may be
regarded as invariance under complex conjugation together with the anti-podal map; in this case the
real structure is a map which takes $CP_1$ onto itself defined on the above functions as
$\eta(\rho) \rightarrow \bar{
\eta(-1/{\bar\rho})}$. The origin of these coordinates and how the Legendre transform is explicitly
constructed out of them has been described in [\use\TwistLT]. We will return to a twistor formulation
of the metrics considered in this paper in the future [\use\progress]. 

The K\"ahler potential is locally constructed in terms of general functions depending on the
holomorphic ${\cal O}(2n)$ coordinates. Given the appropriate contour we define the integral, 

$$ F = {1\over 2\pi i} \oint d\rho ~{1\over\rho^2} G\bigl( 
\eta^{(2n)}(\rho), \rho \bigr) \ .
$$ Note further that by construction this function satisfies many linear differential equations of the
form, 

$$ {\partial^2\over \partial w_i \partial w_j} F  = {\partial^2\over \partial w_{i+a} \partial
w_{j-a}} F \ .
\anoneqn
$$

The K\"ahler potential $K$ is found by performing a complex Legendre transform of $F$ with respect to
$w_1$ and
$w_{2j-1}$. In addition we extremize $F$ with respect to the remaining coefficients other than
$w_0$ or $w_{2n}$. Define $z^i = w_0^i$ (hence ${\bar z}^i = (-1)^n w^i_{2n}$) and $v^i=w^i_1$ (${\bar
v}^i= -(-1)^n w_{2n-1}^i$). Then, the K\"ahler potential is

$$ K(z^i,{\bar z}^i,u^i,{\bar u}^i) = F(z^i,{\bar z}^i,  v^i,{\bar v}^i, w_j) = F - u^i v_i - {\bar
u}^i {\bar v_i}
$$
$$ F_{v^i} = u^i \quad F_{w^j} =0 \ .
\eqn\extreme
$$ The coefficients $w_j$ have indices $(2\leq j\leq 2n-2)$ for each of the $\eta^{(2n)}$ functions.
Finding the metric then involves as usual taking derivatives of the K\"ahler potential.

\vskip .3in
\noindent{\it Case of Maximal Number of Tri-Holomorphic Isometries} \vskip .3in

In the special case in which only a set of ${\cal O}(2)$ type of coordinates $\eta_i$ are necessary an
explicit solution to the K\"ahler potential may be found. We explicitly state the construction for
this case and prove the conjectured form of the ``distinct'' multi-monopole moduli space. The
coordinates and generating function for the local construction of the K\"ahler potential is,

$$
\eta_i = z_i + x_i \rho - {\bar z}_i \rho^2 \quad\quad F= {1\over 2\pi i} \oint d\rho~ G
\bigl(\eta_j (\rho) ,
\rho\bigr) \ . \anoneqn
$$ The standard Legendre transformation with respect to all $n$ coordinates $x_j$ gives a K\"ahler
potential with
$n$ tri-holomorphic isometries. We get from the Legendre transform of the $F$ function

$$
\eqalign{ & K(z_j, {\bar z}_j, u_j, {\bar u}_j) = F(x_j,z_j,{\bar z}_j)  - (u_i + {\bar u}_i) x^i
\cr & {\partial F\over \partial x^j} = u^j+{\bar u}^j \ , }
\eqn\maxmetric
$$ where the $U(1)$ rotations between $u_j$ and ${\bar u}_j$ generate the commuting isometries.

We first write down the most general $4n$-dimensional hyperk\"ahler metric with a maximal number of
tri-holomorphic isometries. In terms of the holomorphic coordinates $u_i, {\bar u}_i$ and $z_i, {\bar
z}_i$ the K\"ahler potential gives the line element as usual from the derivatives,

$$ g = K_{u_i {\bar u}_j} du^i \otimes d{\bar u}^j +  K_{u_i {\bar z}_j} du^i \otimes d{\bar z}^j +
K_{z_i {\bar u}_j} dz^i \otimes d{\bar u}^j + K_{z_i {\bar z}_j} dz^i \otimes d{\bar z}^j \ .
\eqn\genmetric
$$ The solution to the Legendre transform may be found explicitly in the general case by switching to
a non-holomorphic coordinate basis consisting of $x^i, y^i, z^i, {\bar z}^i$ [\use\HkQC,\use\GLT]. The
new coordinates are defined by

$$ 2 du^i = F_{x_i x_j} dx^j + F_{x^i z^j} dz^j + F_{x^i {\bar z}^j} d{\bar z}^j  + i dy^i \ .
\anoneqn
$$ The most general solution to the components of the hyperk\"ahler metric are 

$$ K_{u^i{\bar u}^j} = - (F_{x^j x^i})^{-1} \quad\quad K_{u^i{\bar z}^j} =  (F_{x^k x^i})^{-1} F_{x^k
{\bar z}^j}
$$
$$ K_{z^i {\bar z}^j} = - F_{x^i x^j} - F_{z^i x^k} (F_{x^l x^k})^{-1}  F_{x^l {\bar z}^j}
\anoneqn
$$ Note that these solutions contain both complete and incomplete metrics, and further analysis is
necessary to restrict the form of $F$ to obtain complete ones. Unfortunately there is no canonical
procedure to select the form of $F$ so that it gives only complete metrics (apart from just looking
case-wise for its potential singularities) However, we will prove below that only one such metric
exists in a given $4n$ dimensions which are geodesically complete and satisfy the asymptotic boundary
conditions calculated from the widely separated distinct monopoles. This metric describes the moduli
space of the charge $n$ distinct multi-monopole configuration.

A particularly simple form for the metric may be found by defining new coordinates, the position
variables $2z^j = r_1^j + i r_2^j$, $2{\bar z}^j = r_1^j - i r_2^j$, and $x^j = r_3^j$. After a bit of
algebra one can show that in this basis the metric in eqn.(\use\genmetric) takes the form

$$ ds^2 = M_{ij} d\vec{r}^i \c d\vec{r}^j - \left[ M^{-1} \right]_{ij} q^i q^j 
\quad\quad M_{ij} = -{1\over 2} F_{x^i x^j} \ , \eqn\nongenmetric
$$ with the charge coordinates given by

$$ q^j = i dy^j - F_{z^k x^i} dz^k + F_{{\bar z}^k x^i} d{\bar z}^k \ . \anoneqn
$$ Note that in this coordinate basis the space has generically the commuting isometries $U(1)^n$
generated by constant shifts of the $y_j$ coordinates ($F$ is $y_j$-independent). The existence of
further isometries depends on the precise form of the $F$ functions. An alternative description of the
metrics in eqn.~(\use\maxmetric) has been given in [\use\PapTown].

The multi-monopole moduli space for $n$ distinct fundamental monopoles has $n$ commuting isometries 
[\use\WeinBigGroup ].   Hyperk\"ahler $4n$ dimensional spaces with $n$ tri-holomorphic isometries may
all be locally constructed in the Legendre transform [\use\HkQC]. We expect to find a description of
$M_{\vec{k}}$ in terms of eqn.(\use\nongenmetric) also since the moduli space is known to be
constructible from $n$ of the $\eta^{(2)}$ coordinates (discussed in section IV). From analyzing the
asymptotic forces between embedded fundamental monopole solutions, the asymptotic form of the metric
for $n$ distinct interacting monopoles has been conjectured to be the correct one all the way into the
core [\use\WeinBigGroup]. Define the magnetic charge $g={4\pi/e}$ and $m_i = g
\vec{h}\c \vec{\alpha}^\ast_i$ the mass of the $i^{\rm th}$ fundamental monopole. In this case, the
proposed metric corresponding to $M_{\vec{k}}$ is given in eqn.(\use\nongenmetric) and is that of a
higher-dimensional analog to a Taub-Nut space. The spatial components are given by

$$ 
M_{ij} = {g^2\over 4\pi} {\vec{\alpha}^\ast_i \c \vec{\alpha}^\ast_j \over 
\vert \vec{r}_i - \vec{r}_j \vert} \quad\quad i\neq j 
$$
$$ 
M_{ii} = m_i - {g^2\over 4\pi} \sum_{i\neq j} {\vec{\alpha}^\ast_i \c 
\vec{\alpha}^\ast_j \over \vert \vec{r}_i - \vec{r}_j \vert} \ . 
\eqn\orthometric
$$ 
The matrix $M$ is independent of the $n$ coordinates $y_j$. Note that we sum over all pairs
$(i,j)$; but as noted in [\use\WeinBigGroup], only $n-1$ of the products of simple roots for
$i\neq j$ are non-vanishing for the case of $SU(n)$. In this case one may choose a labelling of the
roots in which only $\vec{\alpha}_i \c
\vec{\alpha}_{i+1} \neq 0$. Then the appropriate center-of-mass variables are the relative distances
$\vec{r}_{i,i+1} = \vec{r}_i - \vec{r}_{i+1}$. However, one may choose any $n-1$ independent linear
combinations of the position coordinates together with the center of mass $\vec{r}=
\sum_{j=1}^n (m_j/m) \vec{r}_j$ when factoring out the flat space to obtain the centered moduli space
$\tilde{\cal M}_{\vec{k}}$.

The charge component $q_i q_j$ of the metric is matched to the previous discussion by performing a
coordinate transformation $i y^j = i w^j + H^j (x,z,{\bar z})$ so that
$$ q^j = idw^j + i H^j_{x^i} dx^i + (iH_{z^i}^j - F_{z^i x^j}) dz^j  + (iH_{{\bar z}^i}^j - F_{{\bar
z}^i x^j}) d{\bar z}^j \anoneqn
$$ One can make the identification with the form of the charge coordinate forces between
well-separated monopoles where 

$$
\eqalign{ q^j & = d\psi^j + \vec{W}^{jk} d\vec{r}^k \cr &
\vec{W}^{jj} = - \sum_{k\neq j} {\vec{\alpha}^\ast_j \c \vec{\alpha}^\ast_k} ~\vec{w}^{jk}
\quad\quad
\vec{W}^{jk} = \vec{\alpha}^\ast_j \c \vec{\alpha}^\ast_k  ~ \vec{w}^{jk} \ . }
\eqn\chargecoord
$$ In eqn.(\use\chargecoord) $\vec{w}^{jk}$ is the Dirac potential from the $k^{\rm th}$ monopole
evaluated at the position $\vec{r}_j$, where the $j^{\rm th}$ Dirac monopole is situated. It satisfies
$\vec{\nabla}_j \times
\vec{w}_{jk}(\vec{r}_j - \vec{r}_k) = - (\vec{r}_j - \vec{r}_k)/ r_{jk}^3$.

\medskip
\noindent{\it Charge Two Cases}
\medskip

The charge two moduli spaces are special in that the possible {\it complete} candidate hyperk\"ahler
metrics with
$SU(2)$ or $SO(3)$ isometry are one of three possible forms (excluding flat $R^4$) [\use\AHSoln,
\use\AHBook]. We briefly discuss this case as its form in the Legendre transform serves to be the
fundamental building block for the higher charge moduli spaces considered in this paper. 

In the case of an
$SU(n>2)$ theory completely broken, there are only two types of charge two monopole moduli spaces. In
the first case, we have both units of magnetic charge aligned in the same direction. The corresponding
moduli space, which was first discovered by Atiyah and Hitchin, has an $SO(3)$ isometry. The latter
case, in which the two charge units live along different weight directions, has a rotational $SU(2)$
in addition to a $U(1)$. The latter factor comes from the additional freedom to gauge rotate along the
Cartan direction orthogonal to total charge conservation (i.e. both directions have separately
conserved electric charges).

Upon imposing an additional $U(1)$ isometry, Atiyah and Hitchin also proved that there is only one
candidate, the Taub-NUT space with negative mass parameter given in eqn.(\use\orthometric) for
$n=1$ [\use\AHSoln]. This classification was used recently to find the moduli space of two distinct
fundamental monopoles [\use\WeinBigGroup,\use\GauntlettBigGroup,\use\Connell]. We have the complete
form of a Taub-NUT metric (with negative mass parameter since $\vec{\alpha}^\ast_1 \c
\vec{\alpha}^\ast_2 <0$) which describes the centered moduli space,

$$ ds^2 = \gamma d\vec{r} \c d\vec{r} + {1\over \gamma} q_r q_r \anoneqn
$$
$$
\gamma = m - {g^2 \over 4\pi} {\vec{\alpha}^\ast_1 \c 
\vec{\alpha}^\ast_2 \over r} \ .
\anoneqn
$$ The form of the Taub-NUT metric is of the type in eq.(\use\nongenmetric) with the generating
function $F$ given by 

$$ F(\vec{r}_i) = {m\over 4 \pi i} \oint_{0} d\rho {\eta^2\over \rho^3}  + {1\over 2\pi i}
\oint_{C} {1\over
\rho^2} 
\eta \{ \ln(\eta) -1\} \ .
\eqn\TNexample
$$ The contour $C$ is taken in a figure-eight fashion around the two zeroes of $\eta(\rho)=0$, and
will be discussed further below. For the distinct moduli space the metric has no singularities and is
geodesically complete. Here the center of mass coordinates $\vec{r}$ and $q$ describe the relative
distance and charge. 

The Taub-NUT metric has either a positive or negative mass parameter depending upon the sign of the
product
$\vec{\alpha}^\ast_1 \c \vec{\alpha}^\ast_2$. With $\vec{\alpha}^\ast_1\c\vec{\alpha}^\ast_2 <0$ we
obtain the complete metric which uniquely describes the moduli space of two distinct monopoles. In the
case of a positive sign we obtain an approximation to the Atiyah-Hitchin metric in the region where
the relative coordinate is very large (compared with the mass). In this case both asymptotic monopoles
are labelled with the same dual simple root (e.g.
$\vec{\alpha}^\ast$), and the metric develops a singularity within the ``core'' region at $r=
{g^2\over 4\pi m}
\vec{\alpha}^\ast_1 \c \vec{\alpha}^\ast_1$. 

\bigskip
\noindent{\it Legendre Description of Distinct Moduli Space} \medskip

The K\"ahler potentials to the distinct multi-monopole moduli spaces considered so far have a
straightforward representation in terms of the generating functions $F$ and are a multi-dimensional
generalization of the Taub-NUT example in eqn.(\use\TNexample). The uncentered higher-dimensional
Taub-NUT analog in eqn.(\use\orthometric) is derived from super-imposing the two fundamental functions
used in writing the Taub-NUT generating function, and is given by
$F=F_{\rm kin} + F_{\rm int}$ where,

$$ F_{\rm kin} = -\sum_{i=1}^n m_i F_1 (\vec{r}_i) \quad\quad F_{\rm int} = \sum_{i\neq j} {g^2\over
4\pi} {\vec{\alpha}^\ast_i \c 
\vec{\alpha}^\ast_j} F_2 (\vec{r}_{ij}) \ . \anoneqn
$$ In terms of ${\cal O}(2)$ coordinates $\eta_i$ these functions are defined by the contour integrals

$$
\eta_i = z_i + x_i \rho - {\bar z}_i \rho^2 $$
$$ F_1(\vec{r}_i) = -{1\over 4 \pi i} \oint_{C_i} d\rho {\eta_i^2\over \rho^3} \quad\quad
F_2(\vec{r}_{ij}) = {1\over 2\pi i} \oint_{C_{ij}} {1\over \rho^2}  (\eta_i-\eta_j) \{
\ln(\eta_i-\eta_j) -1\} \ . \eqn\uncentered
$$ The contours $C_i$ are defined by a small circle (anti-clockwise) around the point $\rho =0$. The
contour
$C_{ij}$ is defined as a figure eight around the two zeroes of $\eta_i - \eta_j =0$ and is illustrated
in fig.(1).

\vskip -.3 cm
\LoadFigure\OTwoContour{\baselineskip 13 pt \noindent\narrower\ninerm The contour $C_{ij}$. The points
$\rho_\pm = -{1/2 z_{ij}} (x_{ij} \pm r_{ij})$ are zeroes of the quadratic equation $\eta_i - \eta_j =
0$.} {\epsfxsize 4.0 truein \epsfysize 1.5 truein}{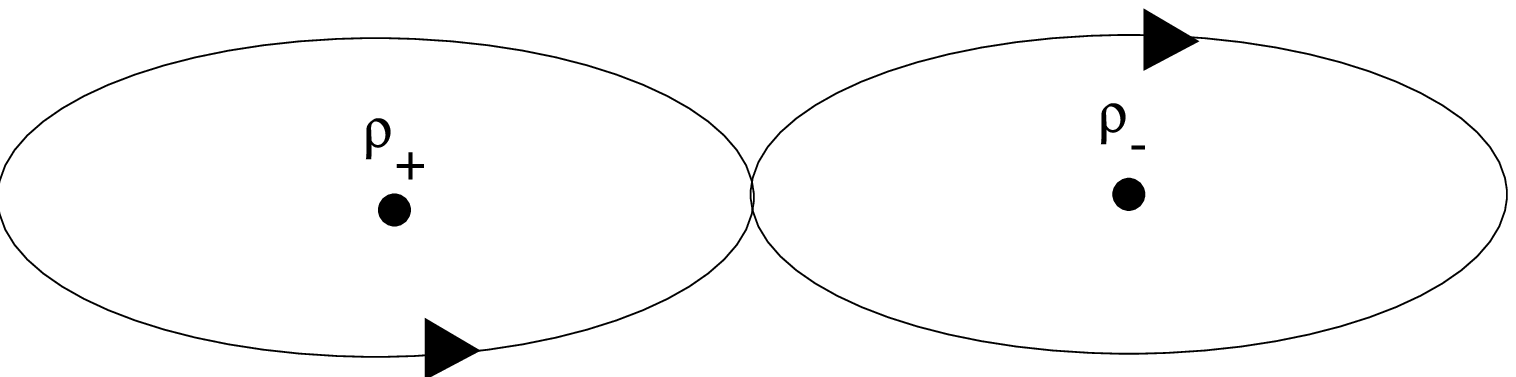}{} \vskip .2in

The flat center-of-mass coordinate is also of the ${\cal O}(2)$ type which may be factored out by
defining a set of
$n-1$ relative coordinates $\vec{{\tilde r}}_i = \vec{r}_i -\vec{r}_{i+1}$. In the ${\cal O}(2)$
coordinate language the new $\tilde{\eta_i}$ are related to the previous ones by $\tilde{\eta_i} =
\eta_i - \eta_{i+1}$. We reproduce the explicit form of the centered Taub-Nut analog here. We first
introduce the $O(2)$ notation corresponding to the distances $\vec{r}_{ij}$ for $i,j = 1,\dots,n-1$.

$$
\vec{r}_{ij} = \vec{{\tilde r}}_{i} + \vec{{\tilde r}}_{i+1} + \ldots + 
\vec{{\tilde r}}_{j-1} \quad\Rightarrow\quad \eta_{ij} = {\tilde \eta}_i + \ldots + {\tilde
\eta}_{j-1} \anoneqn
$$ The metric generated by eqn.(\use\uncentered) with the center of mass factored out is explicitly
generated in the Legendre transform by the function $F=F_{\rm kin} + F_{\rm int}$, 

$$ F_{\rm kin} = - \sum_{i\leq j}^{n-1} m_{ij} F_1 (\vec{r}_{ij}) \quad\quad F_{\rm int} = {g^2\over
2\pi}
\sum_{i<j}^n \vec{\alpha}^\ast_i 
\c \vec{\alpha}^\ast_j F_2(\vec{r}_{ij}) \ , \eqn\centered
$$ where $m_{ij}$ is a general reduced mass matrix found by extracting the center of mass coordinate,

$$
\sum_{j=1}^n m_i \eta^2_i = (\sum_{i=1}^n m_i)  ({\sum_{j=1}^n m_j \eta_j \over \sum_{j=1}^n m_j})^2 +
\sum_{i<j} m_{ij} (\eta_i - \eta_j)^2 \ .
\anoneqn
$$ In the case where the individual masses are constant, $m_i = m$, the matrix is $m_{ij} = (m/n)$. An
evaluation of the $F_i(\vec{r}_i)$ integrals gives

$$ F_1(x,z,{\bar z}) = z {\bar z} - {1\over 2} x^2 $$
$$ F_2(x,z,{\bar z}) = r +x\ln(r-x) - {1\over 2}x \ln(4z{\bar z}) \quad\quad r^2 = x^2 + 4z{\bar z} \ .
\eqn\Ffunctions
$$ One may verify that these functions do indeed generate the uncentered moduli space (as well as its
centered form) given by eqn.(\use\orthometric). 

\bigskip
\noindent {\it Proof of ``Distinct'' Monopole Moduli Space} \medskip

We now prove that the form for $F$ in eqn.(\use\uncentered) is the unique complete hyperk\"ahler
manifold with the maximal number of tri-holomorphic isometries which gives rise to the asymptotic form
of the centered metric found from eqn.(\use\orthometric). The net isometry group is
$U(1)^{n}\times SU(2)$. This then proves the previously conjectured form for the moduli space
describing a set of distinct monopoles to be correct. We do this in two steps. We first note that all
possible $4$-dimensional $\eta^{(2)}$ slices have to be asymptotically locally flat (ALF), complete
with a single $U(1)$ isometry, and possessing a finite set of removable singularities. The asymptotic
form of the metric demands the first condition and tells us the leading behavior; the fact that we are
taking a submanifold composed of an $\eta^{(2)}$ coordinate gives the second condition. The possible
singularities will be discussed below. Given these conditions, the unique form for $F$ is found by
writing down the most general function which has all slices of this type and which satisfies the
asymptotic form of the metric. 

We consider first only four dimensional manifolds. The generating function in the Legendre function
satisfies the Laplace's equation in $R^3$,

$$ F_{xx}+F_{z{\bar z}} = \nabla^2 F(\vec{r}) = 0 \eqn\fequation
$$ We also know the asymptotic form of $F$ from any $4$-dimensional slice of the general
$4n$-dimensional hyperkahler manifold. This is discussed in the section on the long-range interaction
of magnetic monopoles. 

Recently it has been shown that for the distinct monopole moduli space considered here, there are a
finite number of NUT type singularities where the metric diverges as $1/\vert \vec{r}_i -\vec{r}_j
\vert$ [\use\WeinBigGroup]. Four dimensional slices of this larger manifold will then contain in
general a finite number of these singularities.

By differentiating eqn.(\use\fequation) one obtains $\nabla^2 F_{xx} = 0$. However, this is not
entirely true since there are non-trivial removable singularities. In general the $F_{xx}$ functions
satisfy the three dimensional Laplace's equation everywhere but at a finite number of points where the
metric possess the NUT singularities. The spatial components of the metric are obtained from $F_{xx}$,
and at a finite number of points $\vec{b}_j$ it behaves as $1/\vert
\vec{r}-\vec{b}_j \vert$. Then we have in terms of the Dirac delta function,

$$
\nabla^2 F_{xx} = \sum_{j=1}^{n} c_j \delta(\vec{r} - \vec{b}_j) \ . \eqn\fxxequation
$$

Finding the solution to eqn.(\use\fxxequation) is a straightforward electrostatics problem. From the
asymptotic locally flat condition, so that $F_{xx}$ approaches a positive constant $a$ at infinity,
the general solution, which we write in terms of the $F$ function, is 

$$ F = - a \sum_{j=1}^{m} F_{1}(\vec{r}-\vec{a}_j) + 
\sum_{j=1}^{n} {c_j\over 4\pi} F_2 (\vec{r} - \vec{b}_j ) \ . \anoneqn
$$ The functions $F_1$ and $F_2$ are defined in eqns.(\use\Ffunctions). These metrics belong to the
multi-Taub-NUT class with a finite number of mass points $\vec{b}_j$ (for $a\neq 0$). Demanding
completeness fixes the signs of the coefficients $c_j$ to all be {\it negative} and $a$ {\it positive}
(or both with opposite signs).

Recall that the distinct $n$-monopole moduli space is described by $n$ of the ${\cal O}(2)$
coordinates. We first take a generic $U(1)$ invariant $4$-dimensional slice of the
$4n$-dimensional distinct moduli space; the submanifold described by one coordinate $\eta$ is found by
fixing all $\eta_i$ coordinates except for one linear combination $\eta$. The space must be ALF and
thus given by a superposition of mass point interactions, $F = F_{\rm kin} + F_{\rm int}+ F_{\rm con}$,

$$ F_{\rm kin} = \sum_{j=1}^r \kappa_j F_1(\vec{r}-\vec{a}_j) \quad\quad F_{\rm int} =
\sum_{j=1}^s \lambda_j F_2(\vec{r}-\vec{b}_j) \ . \eqn\oneslice
$$ The remaining $F_{\rm con}$ is a constant from the viewpoint of $\eta$ (i.e. may depend on the
remaining independent combinations of $\eta_i$). In eqn.(\use\oneslice), the parameters $a_j$,
$b_j$, $\kappa_j$, and
$\lambda_j$ are $\eta$-independent, but may in general depend on the remaining frozen ${\cal O}(2)$
coordinates. Note that if all $a_j$ and $b_j$ were the same the metric would acquire an additional
$SU(2)$ isometry and then give us a Taub-NUT space. The general form in eqn.(\use\oneslice) resembles
a multi-Taub-NUT space. 

All generating functions $F$ for the generic $4$-dimensional slices defined by fixing all $\eta_j$
coordinates except for {\it any} one linear combination must have the same functional form as in
eqn.(\use\oneslice). The fact that any of the $\eta$-type submanifolds obey this property is a very
stringent condition on the entire generating function (or rather, the K\"ahler potential) of the
$4n$-dimensional metric. 

The form of $F$ in eqn.(\use\oneslice) demands a functional linear dependence in $\eta$, both within
the logarithm and square, in the $4n$-dimensional result containing the entire contribution
$F_{\rm int}$ and $F_{\rm kin}$. For ease of notation define a finite set of general linear
combinations of the $\eta_i$ coordinates through the matrices $M$ and $N$ as 

$$ {\tilde\eta}_p = M_{pi} \eta_i \quad\quad p ~{\rm finite} $$
$$ {\hat\eta}_q = N_{qi} \eta_i \quad\quad q ~{\rm finite} $$ The unique functional form for the
generating function with this slice independent linear property is 

$$
\eqalign{ {\tilde F} = & {1\over 2\pi i} \sum_p \alpha_p 
\oint_{{\tilde C}_p} d\rho~ {\tilde\eta}_p^2 {1\over \rho^3} \cr & + {1\over 2\pi i} \sum_q
\beta_q \oint_{{\hat C}_q}  d\rho~ {1\over \rho^2} ({\hat\eta}_q)
\{ \ln({\hat\eta}_q) -1 \} \ . }
\eqn\genform
$$ The coeffients $\alpha$, $\beta$, $M$, $N$ are arbitrary constants which do not depend on the
coordinates. We demand next that ${\tilde F}$ has agreement with the known asymptotic form of the
monopole moduli space, is geodesically complete, and invariant under overall translations $\eta_j
\rightarrow \eta_j + {\bar\eta}$ for all
$j$ simultaneously. These conditions uniquely fix the coefficients in eqn.(\use\genform) so that we
have agreement with the centered form in eqn.(\use\centered) with the negativity condition on the
coefficients
$\vec{\alpha}_i^\ast \c \vec{\alpha}_j^\ast$. This proves that the previously conjectured form of the
moduli space metric for distinct monopoles is in fact the correct one.

\vskip .4in
\noindent{\bf IV. Aligned Multi-Monopoles} \vskip .3in

The multi-monopole configuration for distinct fundamental monopoles is special in that there is no
charge exchange which modifies the ``core'' of the moduli space. The general $SU(2)\rightarrow U(1)$
uncentered multi-monopole moduli space is described in the generalized Legendre transform construction
by the configuration of coordinates
$\oplus_{j=1}^k \eta^{(2j)}$ (coming from an intermediate projection of the twistor space $Z_k
\rightarrow
\oplus_{j=1}^k {\cal O}(2j) \rightarrow CP_1$) [\use\AHBook]. 

The multi-monopole moduli space for $k$ distinct monopoles in $SU(n)$ and the configuration where two
of the monopoles live along the same root directions are described furthermore by the sets of
coordinates $\oplus_{j=1}^k
\eta^{(2)}$, $\eta^{(2)} \oplus \eta^{(4)} \oplus_{j=1}^{k-1} \eta^{(2)}$ respectively
[\use\projections]. In general, the types of coordinates needed are given by the $SU(2)$ number for a
charge $k_i$ monopole along each of the components of the total magnetic charge vector. Note that
charge configurations with $k_i \leq 2$ describe ``multi-Atiyah-Hitchin'' type of metrics and only
require the $\eta^{(2)}$ and $\eta^{(4)}$ coordinates within the generalized Legendre transform.
Furthermore, the general monopole will always have at least one ${\cal O}(2)$ type of coordinate
serving as the center of mass which is factored out to obtain the centered moduli space.

The presence of the section $\eta^{(4)}$ in the description of the aligned multi-monopole moduli space
considerably complicates the construction of the moduli space metrics, for one has to solve a series
of constraints (extremizing the $F$ function with respect to the coefficients $w_2$ in
eqn.(\use\extreme)) in the generalized Legendre transform. However, even without solving explicitly
these constraints one may find the asymptotic form of the metric which agrees with the one from
well-separated fundamental monopoles.

Recently the Atiyah-Hitchin metric has been worked out within the generalized Legendre transform
construction [\use\TwistLT]. In this example the centered $2$-monopole moduli space is described by an
${\cal O}(4)$ coordinate
$\eta$ and generating function, 

$$ F_{AH} = -{m\over 2\pi i} \oint_C d\rho~ {\eta\over \rho^3} +  {g^2 \vec{\alpha}^\ast_1 \c
\vec{\alpha}^\ast_1
\over 4\pi}  {1\over 2\pi i} \oint_{\tilde C} d\rho~ {1\over \rho^2} \sqrt{\eta} \ .
\eqn\AHfunction
$$ The asymptotic regime of the moduli space approaches a Taub-NUT space with positive mass parameter,
and one expects a limiting form of the $F$-function to approach the description in
eqn.(\use\TNexample) but with
$\vec{\alpha}^\ast_1 \c\vec{\alpha}^\ast_1$ replaced with $\vec{\alpha}^\ast_1
\c\vec{\alpha}^\ast_2 >0$. The asymptotic limit on the level of the $F$ function can be understood as
a degeneration of the ${\cal O}(4)$ coordinate into a square of an ${\cal O}(2)$ one, $\eta
\rightarrow (\eta^{(2)})^2$. This limit will be described in detail below, but the functional form of
$F_{AH}$ in eqn.(\use\AHfunction) is important as it provides an interesting way to generalize the
distinct monopole moduli spaces to those containing two like-charge components. 

Our results generalize the Atiyah-Hitchin metric by including additional charges along different
$U(1)$ directions. An example of a charge vector within a broken $SU(n)$ theory we analyze here is

$$
\vec{q}_m = g \bigl( 2\vec{\alpha}^\ast_1 + \vec{\alpha}^\ast_2 + \ldots + \vec{\alpha}^\ast_p
\bigr)
\eqn\multicharge
$$ for $p\leq n$.

In the following we take the center of mass frame corresponding to the positions one and two (i.e.,
$\vec{r}_1+\vec{r}_2=0$). In the twistor picture we required the types of coordinates $\eta^{(2)}$ and
$\eta^{(4)}$ from the charge two component. This frame choice means that this particular
$\eta^{(2)}$ coordinate is the factored out center of mass coordinate.

Our $F$-function conjectre is found by ``enlarging'' one of the $\eta^{(2)}$ coordinates used in
formulating the distinct moduli space with an $\eta^{(4)}$ one. The motivation behind this is that in
the asymptotic regimes these larger coordinates behave like the square of an $\eta^{(2)}$, as will be
shown. We conjecture that the form of the monopole moduli space corresponding to the charge in
eqn.(\use\multicharge) is correct {\it after} replacing the relative coordinate describing
$\vec{r}_1-\vec{r}_2$ in the asymptotic form with an $\eta^{(4)}$ one. 

In the following we simplify the notation by identifying the two vectors $\vec{\alpha}^\ast_1
=\vec{\alpha}^\ast_2$ and taking all mass parameters to be the same. Thus we consider the
multi-monopole configuration which asymptotically corresponds to the charge configuration given by
$\vec{q}_m = g \sum_{j=1}^p \vec{\alpha}^\ast_j$ with the $j=1,2$ identification. Then our proposed
$F$ function is 

$$ 	
\eqalign{ & F_{\rm kin} = -{1\over 4\pi i}
\oint_{C} d\rho ~ {1\over \rho^3} \bigl( {{\eta\over 2} +\sum_{i=2}^{p} \eta^2_i} \bigr)
\cr & F_{\rm int} = \sum_{i<j}^{p} {g^2\over 4\pi} \vec{\alpha}^\ast_i \c 
\vec{\alpha}^\ast_j F_{\rm int}^{ij} \ , }
$$ where the $F^{ij}$ generating functions are defined by $$
\eqalign{ & F^{12} = {1\over 2\pi i} \oint_{{\tilde C}_{12}}  d\rho ~{1\over \rho^2} \sqrt{\eta}
\cr & F^{ij} = {1\over 2\pi i}~ \oint_{{\tilde C}_{ij}} d\rho~  {1\over \rho^2} \eta_{ij}
(\ln{\eta_{ij}} -1)
\quad\quad (i,j)\neq (1,2)
\ . }
\eqn\alignedmod
$$ The coordinates in eqn.(\use\alignedmod) are given by 

$$
\eqalign{ &
\eta_{ij} = \eta_i - \eta_j \quad\quad (i,j>2) \cr &
\eta_{1j} = \eta_j + {\sqrt{\eta}\over 2} \quad\quad (j>2) \cr &
\eta_{2j} = \eta_j - {\sqrt{\eta}\over 2} \quad\quad (j>2) \ . }
\eqn\newcoords
$$ Of course, we must be very careful in defining the contours ${\tilde C}_{ij}$ in the above
integrals. There are three types of integrals appearing in eqn.(\use\alignedmod).

The contours are chosen as follows. $C$ is a closed curve encircling the origin once in the
counter-clockwise direction. The remaining contours ${\tilde C}_{ij}$ are either from integrals
containing only ${\cal O}(2)$ coordinates (latter in eqn.(\use\newcoords)) or with only one ${\cal
O}(4)$ (former). The completely ${\cal O}(2)$ integrand is taken with the previously described
Taub-NUT contour; ${\tilde C}_{ij}$ is defined in a figure eight fashion around the two zeroes of
$\eta_{ij}=0$. 

The mixed integral is more complicated since the integrand is single-valued on a double cover of the
logarithmic Reimann surface. However, keeping in mind that we want the correct asymptotic limits of
the moduli space limits severely the choice of contour. The integrals are described and evaluated in
Appendix I. In the following we investigate the asymptotic limits of the moduli space and show that it
has the correct form for the multi-monopole moduli space described by the charge configuration in
eqn.(\use\multicharge).

\bigskip
\noindent{\it Kinetic Contribution to $F_w=0$ Constraint} \bigskip

We first evaluate the $w$-equation constraint on the form for $F$ in eqn.(\use\alignedmod). The
kinetic contribution splits into two types of integrals (${\cal O}(2)$ and ${\cal O}(4)$), $$
\eqalign{ F_{\rm kin}^{j} & = {1\over 2\pi i} \oint_0 d\rho~ {\eta_{j}^2\over \rho^3} \cr & = -{1\over
2} x_{j}^2 + z_{j} {\bar z}_{j} }
\quad\quad\quad
\eqalign{ F_{\rm kin}^{1} & = -{2 m \over 2\pi i} \oint_0 {\eta\over 2 \rho^3} \cr & = -m w \ . }
\anoneqn
$$ Furthermore, the contribution to the $w$-constraint from the three types of ``kinetic'' terms above
is very simple, 

$$ F_{{\rm kin}, w}= \partial_w F_{\rm kin} = -m \ . \anoneqn
$$

\bigskip
\noindent{\it Interaction Contribution}
\bigskip

Next we list the contributions from the interaction part of the prepotential. The Taub-NUT type of
interaction from orthogonally charged monopoles is given from the terms with $i,j>2$. 

$$
\eqalign{ F_{\rm int}^{ij} & = {1\over 2\pi i} \oint_{{\tilde C}_{ij}} d\rho ~  {1\over \rho^2}
\eta_{ij} \bigl(
\ln{\eta_{ij}} -1 \bigr) \cr & = F_2 (\vec{r}_{ij}) = r_{ij} + x_{ij} \ln( r_{ij}-x_{ij}) -{1\over 2}
\ln(4 z_{ij} {\bar z}_{ij}) }
\anoneqn
$$ Next, from the two similar monopoles with $i=1,j=2$ we have an Atiyah-Hitchin type of interaction.
The contour integral is evaluated in Appendix 1 and is explicitly,

$$ F_{\rm int}^{12} = {1\over 2\pi i} \oint_{{\tilde C}_{12}} {1\over \rho^2}~ 
\sqrt{\eta}
\anoneqn
$$
$$
\partial_w F_{\rm int}^{12} = {1\over 4\pi i} \oint_{{\tilde C}_{12}} d\rho~  {1\over\sqrt{\eta}} =
{4\over \left[ c (1+ \beta{\bar\beta})(1+\alpha{\bar\alpha})\right]^{1\over 2}}  {\cal K}(k) \ ,
\anoneqn
$$ where ${\cal K}$ is a complete integral of the first kind, and the (real) modulus is $$ k^2 =
{(1+{\bar\alpha}\beta)(1+\alpha{\bar\beta})\over  (1+\beta{\bar\beta})(1+\alpha{\bar\alpha})} \ .
\anoneqn
$$ The final type of integral comes from the pairs $i=1,j>2$ and $i=2,j>2$. In this case we obtain a
contribution which is free of any ambiguity in defining the square root of the ${\cal O}(4)$
coordinate,

$$
\eqalign{ F_{{\rm int},w}^{1j} + F_{{\rm int},w}^{2j} = &  {1\over 8\pi i} \oint_{{\tilde C}_{1j}}
{1\over
\sqrt{\eta}} \bigl\{ \ln{ (\eta_{j} + {\sqrt{\eta}\over 2})} + 1 \bigr\} \cr & - {1\over 8\pi i}
\oint_{{\tilde C}_{2j}} {1\over \sqrt{\eta}} \bigl\{ \ln{ (\eta_{j} - {\sqrt{\eta}\over 2})} + 1
\bigr\} }
\anoneqn
$$
\medskip
\noindent These integrals have been evaluated in Appendix I and we state the complete result below.

The final form for the constraint imposed by requiring the $w$-derivative to vanish is then

$$
\eqalign{ F_w & = 0 = -m + {g^2 \over \pi} {1\over\left[ c (1+ \beta{\bar\beta}) 
(1+\alpha{\bar\alpha})\right]^{1\over 2}} \Bigl\{ 2\vec{\alpha}^\ast_1 \c \vec{\alpha}^\ast_1 {\cal
K}(k)
\cr & + \sum_{j=3}^{n} \vec{\alpha}^\ast_1 \c \vec{ \alpha}^\ast_j 
\Bigl[ {\cal F}[\phi_R(\delta^j_+),q] - {\cal F}[\phi_R(-1/{\bar\delta}^j_-),q] + {\cal
F}[\phi_L(\delta^j_-),q] - {\cal F}[\phi_L(-1/{\bar\delta}^j_+),q] \Bigr] ~\Bigr\} \ . }
\eqn\wconstraint
$$
\medskip
\noindent The integrals above and the notation are defined in Appendix 1. We next must analyze the
form for the K\"ahler potential subject to this constraint to determine the correct asymptotic limits.
Unlike for the solution to the AH metric in the generalized Legendre transform, eqn.(\use\wconstraint)
may not be directly inverted to obtain the overall scale $c$ of the two ``like'' monopole
interactions.

\bigskip
\noindent{\it Asymptotic Limits}
\bigskip

The limits in which we pull apart the two like monopoles or all $n$ simultaneously asymptotically far
apart should reproduce the metric discussed previously in section. The difference will be that in this
case one of the products of dual vectors is less than zero and hence reproduces a Taub-NUT type of
space with positive mass parameter (which does not have a complete metric). The corrections are
suppressed exponentially and resolve the potential singularity, just as in the Atiyah-Hitchin metric.
Furthermore, the generating function has been given in the first two particles' center of mass frame.

In the following we examine eqn.(\use\wconstraint) in the various asymptotic limits and determine the
behavior of the ${\cal O}(4)$ coordinate. Rather than rederiving the metric explicitly in these
regions we find the asymptotic form of the generating function in the Legendre transform.

\bigskip
\noindent{\it $\vec{r}_{12} \rightarrow \infty$ Limit} \bigskip

Consider the limit in which the branch cuts of the elliptic surface defined by $\eta=0$ move on top
one another. In this case $\alpha=\beta=\mu$ and ${\bar\alpha}={\bar\beta}={\bar\mu}$ and the real
modulus of the curve approaches one. This limit has been shown in the derivation of the Atiyah-Hitchin
metric, from solving the $w$-equation up to the scale $c$, to correspond to the regime of separating
the two monopoles far apart, since in this case the overall scale is set by the complete elliptic
integral with modulus $k$. Although the roots have not been solved for explicitly in terms of the
universal scale $c$ from eqn.(\use\wconstraint), one can see that this limit corresponds to
degenerating the ${\cal O}(4)$ coordinate into an ${\cal O}(2)$ type in which an additional
approximate tri-holomorphic $U(1)$ isometry is regained. This is physically what one expects as the
widely separated monopoles have an approximate individual charge conservation.

As the two roots $\alpha$ and $\beta$ come close together, we find that the polynomial $\eta$ limits
to ,

$$
\eqalign{
\eta & = c (\rho-\alpha)(\rho-\beta)(\rho{\bar\alpha}+1)  (\rho{\bar\beta}+1)
\cr &
\rightarrow c(\rho-\mu)^2 (\rho{\bar\mu}+1)^2 = {\bar z} (\rho-\mu)^2  (\rho+{1\over{\bar\mu}})^2
\ , }
\eqn\mulimit
$$ where $\alpha,\beta\rightarrow\mu$. A transformation to new coordinates $z'$, $u'$ then allows the
limit in eqn.(\use\mulimit) to be expressed as 

$$
\eta' \rightarrow {({\bar z}'})^2 (\rho-\mu')^2  (\rho+{1\over{\bar\mu}'})^2 \ ,
\anoneqn
$$ in which we see exactly the square of an ${\cal O}(2)$. Of course we should examine the
$w$-equation explicitly in order to show that this generic behaviour is indeed correct. We will do
this and explicitly give the coordinate transformation. Since we are looking at the location of the
four zeroes, however, we know implicitly the behaviour of $w$ and how this procedure will go through.

The limiting form of the incomplete elliptic integrals is sub-leading. Consider what happens as
$\alpha\rightarrow\beta$ and $c\rightarrow\infty$ to the four zeroes of ${\sqrt{\eta}/2} \pm
{\tilde\eta}=0$ degenerate into two solutions as the coordinates in $\tilde\eta$ are suppressed in
comparison to ${\bar z}^{1/2}$. In fig.(3) we have $\delta_-, -1/{\bar\delta_+} \rightarrow
\alpha$ and $\delta_+, -1/{\bar\delta_-} \rightarrow
\beta$. Each incomplete integral is in fact finite in this limit. The real $q^2$ modulus in
eqn.(\use\wconstraint) approaches zero, 

$$ q^2 = {({\bar\beta}-{\bar\alpha})(\beta-\alpha) \over  (1+{\bar\beta}\beta)(1+{\bar\alpha}\alpha)}
\rightarrow 0
\ . \anoneqn
$$

In the leading limit the difference of the left or right incomplete integrals also approaches zero,

$$
\lim_{(\delta_+,-1/\delta_-) \rightarrow \beta}  {\cal F}[\Phi^R(\delta_+), q] - {\cal
F}[\Phi^R(-1/\delta_-), q] 
\rightarrow 0 \ .
\eqn\lrdiff
$$ The behaviour of eqn.(\use\lrdiff) can be shown to approach $0$ as ${\cal O}(1/\sqrt{\bar z})$. All
contributions from the incomplete integrals are suppressed in comparison to the complete
${\cal K}(k)$ contribution, which diverges logarithmically as $\ln(1-k^2)$ when $k$ goes to one.

We first define the leading divergent term from the complete integral as,

$$
\tau = ({2g^2 \vec{\alpha}^\ast_1 \c \vec{\alpha}^\ast_1 \over m\pi})^2 
\lim_{k\rightarrow 1} {\cal K}^2(k) = ({g^2 \vec{\alpha}^\ast_1 \c \vec{\alpha}^\ast_1 \over 2m\pi})^2
\ln^2(1-k) + {\cal O}(1)
$$ The solution to $c$ in eqn.(\use\wconstraint) as $k\rightarrow 1$ is then to leading order,

$$ c = {1 \over (1+\mu{\bar\mu})^2} \tau + \ldots \quad\quad \ . \eqn\ahlimit
$$ Not surprisingly, the limit in eqn.(\use\ahlimit) gives precisely the solution obtained in solving
for the Atiyah-Hitchin metric. We parametrize the roots of the ${\cal O}(4)$ coordinate in terms of
angular variables
$\Phi$, $\Psi$, and $\theta$, and solve as was done in [\use\TwistLT], for the coefficients of
$\eta$ subject to the constraint from eqn.(\use\ahlimit). In terms of these angular variables, the
$\tau$ dependent relative coordinates are 

$$
\eqalign{ & z = {1\over 4} e^{2i\phi} \Bigl[ \cos^2(\psi) - \sin^2(\psi) \cos^2(\theta)  + i
\sin(2\psi)
\cos(\theta) \Bigr] \tau + \ldots \cr & v = e^{i\phi} \sin(\theta) \sin(\psi) \Bigl[ \cos(\psi) + i 
\cos(\theta) \sin(\psi) \Bigr] \tau + \ldots \cr & w = {1\over 4} \Bigl[ -2 + 6 \sin^2(\theta)
\sin^2(\psi) \Bigr]
\tau \ . }
\eqn\coords
$$ Explicitly we find that the scale factor of the coordinates diverges, so that the monopole ``pair''
is being pulled apart. 

We now rewrite the coordinates in eqn.(\use\coords) in the manner, 

$$
\eqalign{ & z = {z'}^2 = \Bigl\{ {1\over 2} e^{i\phi} \bigl[ \cos(\psi) + i \cos(\theta) 
\sin(\psi) \bigr] \Bigr\}^2 \tau
\cr & v = -2 {z'} x' = - 2 \Bigl\{ {1\over 2} e^{i\phi} \bigl[ \cos(\psi) + i \cos(\theta) 
\sin(\psi) \bigr] \Bigr\} \bigl[ - \sin(\theta)\sin(\psi) \bigr] \tau \ . }
\anoneqn
$$ Then we have the solution to $w$
$$ w = {x'}^2 - 2{z'}{\bar z}'
\anoneqn
$$ In the new coordinates the ${\cal O}(4)$ polynomial may be written as 

$$
\eta = (z' + x' \rho - {\bar z}' \rho^2)^2 + \ldots \anoneqn
$$

After the change of variables given above we find the limiting form of the ${\cal O}(4)$ coordinate to
be
$\eta\rightarrow {\tilde\eta}^2$, where the latter is of the ${\cal O}(2)$ type with the functional
form
${\tilde\eta} = z + x \rho - {\bar z} \rho^2$. Note that this is correct asymptotically only while
taking the
$w$-constraint into account, since it is crucial for $w$ to have a value such that the degree four
polynomial factorizes. The asymptotic limit $\vert \vec{r}_{12} \vert >> m$, of the ${\cal O}(4)$
coordinates into a square of the ${\cal O}(2)$ type is what initially lead to the investigation and
proposals for the classes of moduli spaces considered in this paper.

The crux to understanding these limits is not necessarily to examine all of the non-trivial equations
in the Legendre transform, but to look at the locations of the zeroes of the three types of
polynomials. The coalescing branch points and zeroes determine the location of potential singularities
in the metric and the asymptotic regimes in the coordinate expansions. In the above limit the two
branch cuts moved on top of one another, and in the process have cancelled eachother. In this manner
we regain the features of an ${\cal O}(2)$ type of contribution. 

Thus without explictly dealing with the components of the metric we may find the form in this region.
We substitute in the appropriate form of $\eta$ in the asymptotic limit of $F$ into
eqn.(\use\alignedmod), where in this regime the coordinates are defined by 

$$
\eqalign{ &
\eta_{ij} = \eta_i - \eta_j \quad\quad (i,j>2) \cr &
\eta_{1j} = \eta_j + {\eta\over 2} \quad\quad (i=1) \cr &
\eta_{2j} = \eta_j - {\eta\over 2} \quad\quad (i=2) \ . }
\eqn\coordinates 
$$ 
This gives $F$ and the metric to sub-leading order as $\vec{r}_{12}\rightarrow\infty$. Note that in
the limit of the degeneration of the $O(4)$ coordinate into the square of the 
$O(2)$ one, $\eta^{(4)}\rightarrow (\eta^{(2)})^2$, the contour integral
$F^{12}_{\rm int} = {1\over 2\pi i} \oint_{{\tilde C}_12} d\rho~{1\over
\rho^2} \sqrt{\eta^{(4)}}$ goes over into the function described by the contour integral 
${1\over 2\pi i} \oint_C d\rho~ \eta^{(2)} (\ln \eta^{(2)}-1)$ [\use\TwistLT ].   The  logarithm in
the latter serves to define the choice of contour.  Together with  the replacement  of
$\eta^{(4)}\rightarrow (\eta^{(2)})^2$ in the remaining integrals $F_{1j}$ and 
$F_{2j}$   in eq.(\use\alignedmod) and eq.(\use\coordinates) gives the asymptotic form of the  moduli
space metric for its higher dimensional Taub-NUT analog.  

In summary, the asymptotic metric is of the higher-dimensional Taub-NUT type and agrees with that
calculated from the asymptotic forces. Furthermore, because of the square root coordinate change we
see that the metrics in eqn.(\use\alignedmod) are a double cover asymptotically of the
higher-dimensional Taub-NUT space.

\bigskip
\noindent{\it $\vert\vec{r}_{ij}\vert \rightarrow \infty$ for all $i$,$j$}
\bigskip

Although the previous limit is the most interesting, we will describe another expansion to illustrate
the basic ideas. We consider scaling {\it all} length scales while keeping in general the ratios
fixed. Again, the contribution from the incomplete integrals is suppressed relative to the
contribution ${\cal K}(k)$. In taking the limit $\vert \vec{r}_{ij} \vert >> 1$ for all $i,j$ the
zeroes of the equations $\sqrt{\eta}/2 \pm \eta_j=0$ do not coincide and do not head off to infinity
(they depend on ratios of the large coordinates). As before, the large
$\vec{r}_{12}>>1$ limit requires, however, that $\alpha\rightarrow\beta$. Examining the integral
representation for the ${\cal F}$ functions in eqn.(\use\wconstraint) from the incomplete integral
contribution we see that none of the integrals have a potential divergence, either from poles within
the denominator or from the limits of integration. The same analysis as above goes through and we find
that the asymptotic form of the metric when all monopoles are widely separated leads to the same one
as in the previous large $\vec{r}_{12}$ limit.

\bigskip
\noindent{\it $D_k$ Metrics}
\medskip

It is interesting that the metrics proposed for this monopole moduli space has a strong similarity to
the $D_k$ asymptotically locally euclidean spaces. This (four-dimensional) class of metrics is
believed to be found within the generalized Legendre transform from the generating function
[\use\TwistLT]

$$ G = \sum_{j=1}^k (\sqrt{\eta} + a_j)\ln(\sqrt{\eta}+a_j) +  (\sqrt{\eta} - a_j)\ln(\sqrt{\eta}-a_j)
\ . \anoneqn
$$ The coordinate $\eta$ is of the ${\cal O}(4)$ type and the parameters $a_j$ are ${\cal O}(2)$
constants which parameterize the $D_k$ moduli space.

Returning to the moduli space considered in detail in this paper, the kinetic term of the $F$ function
given in eqn.(\use\alignedmod) with the different mass parameters inserted is actually 

$$ F_{\rm kin} = -{1\over 4\pi i}
\oint_{C} d\rho ~ {1\over \rho^3} \bigl( m_1 {{\eta\over 2} +\sum_{i=2}^{p} m_2 \eta^2_i} \bigr) \ .
\anoneqn
$$ By leaving $m_1\neq 0$ we appear to have an asymptotically locally flat version of the $D_k$ space.

In the limit $m_1=0$ we obtain a ``fluctuating'' asymptotically locally euclidean $D_k$ space. By
fluctuating we mean that by fixing the ${\cal O}(2)$ coordinates $\eta_i$ (i.e. taking different
four-dimensional ${\cal O}(4)$ submanifolds) the $D_k$ space is actually recovered. The $\eta_i$
coordinates correspond to mass point insertions - the moduli of the $D_k$ space. It is unknown at this
point what the interesting connection is between the proposed moduli space and the $D_k$ ALE spaces
(or its ALF variants).

\vskip .4in
\noindent{\bf V. Generalizations and Multi-Atiyah-Hitchin Metrics} \vskip .3in

Starting from the form of the generating function $F$ which describes the distinct moduli space, it is
possible to conjecture entire classes of generating functions which describe multi-Atiyah-Hitchin type
metrics. These spaces have magnetic charges with no more than two units along any dual vector
direction; specifically, the charge vectors are labelled by

$$
\vec{q}_m = \sum_i n_i \vec{\alpha}^\ast_i \quad\quad n_i = 0,1,2 \ . \anoneqn
$$ It is possible to conjecture the form of $F$ which gives these monopole moduli spaces for several
reasons. 

First, the {\it centered} moduli space for a monopole containing $n$ charge-two and $m$ charge-one
components must be described in terms of $n$ of the $\eta^{(4)}$ and $m+n-1$ of the $\eta^{(2)}$
coordinates. These spaces have real dimension $d=4(2n+m-1)$, and in the asymptotic regime where we
pull out the charge-two components one expects a multiple Atiyah-Hitchin type of metric. 

Second, the analysis presented in the previous section shows how generically the $\eta^{(4)}$
coordinates limit into an $\eta^{(2)}$ type as we pull its coordinates out to infinity (or rather, how
its four roots degenerate into two pairs). The form of the generating function in eqn.(\use\centered)
which describes the well-separated multi-monopole configuration, the ``distinct'' case but with
appropriate sign changes coming from the inner products
$\vec{\alpha}^\ast_i \c \vec{\alpha}^\ast_j$, must be satisfied by any conjecture in the asymptotic
regime.

The route we take to conjecture these generating functions is to simply replace the ${\cal O}(2)$
coordinates with the square roots of the ${\cal O}(4)$ ones in the description of the asymptotic form
of the moduli space. For example, the $F$ function describing the charge $q_m = 2\vec{\alpha}^\ast_1 +
2\vec{\alpha}^\ast_2 +
\vec{\alpha}^\ast_3 + \ldots$ is found by replacing $\eta_1$ and $\eta_2$ in eqn.(\use\centered) with
$\sqrt{\eta^{(4)}_1}$ and $\sqrt{\eta^{(4)}_2}$. The contours must also be redefined, the detailed
form of which depends on the number of ${\cal O}(4)$ coordinates but in an analagous fashion to the
case with two units of charge along one component considered previously.

\vskip .4in
\noindent{\bf VI. Discussion}
\vskip .3in

We have examined the multi-monopole moduli spaces in $SU(n)$ gauge groups from the starting point of
the Legendre transform technique and its generalizations. In the case of the moduli space with
$k<n$ $U(1)$ isometries, which describes the monopole with no more than one unit of charge along each
Cartan direction, we have proved the unique form to be an analog of a Taub-NUT space in $4k$
dimensions. Looking at the generating function of the K\"ahler potential from the viewpoint of the
Legendre construction, we have been able to conjecture further examples of moduli spaces in which
there are no more than two units of magnetic charge in a single direction. These examples of metrics
all have the correct asymptotic behavior and the appropriate isometries, although we have not been
able to prove them to be the unique form. They correspond to joining together multiple Atiyah-Hitchin
metrics and Taub-NUT type interactions in a hyperk\"ahler fashion.

There are a number of remaining open questions to be addressed. One would like an explicit form for
the metrics given by the generating functions proposed in this paper, which involves solving a system
of complicated constraint equations. In addition, the question of uniqueness of these examples should
also be addressed. 

The predictions of Montonen-Olive duality in certain supersymmetric theories demand that the spectrum
of elementary vector bosons be consistent with that of the magnetic counterpart. Under an
$SL(2,Z)$ transformation each vector gets mapped to a tower of magnetically charged states, which in
principle may be found by looking at appropriate harmonic forms on the monopole moduli spaces. These
conjectures may be tested further for higher gauge groups if one knew the relevant moduli spaces, the
form for several examples we propose here.

We also find it curious that the metrics proposed here are related to the asymptotically flat versions
of the $D_k$ and $A_k$ class of metrics. It would be interesting to find the physical reason for the
appearance of these types of spaces in the moduli spaces considered here.

\vskip .3in
\noindent{\bf Aknowledgements}
\vskip .3in

The author would like to thank M. Rocek for many valuable discussions during the course of this work
and I. Ivanov for useful discussions about twistor spaces. Useful discussions with Jan DeBoer, Alfred
Goldhaber, and Claude LeBrun are also aknowledged.  This research supported by National Science
Foundation grant PHY-9309888.

\vskip 1.0in
\Appendix{Integrals}

There are three integral forms necessary to describe the proposed ``aligned'' monopole moduli space.
They have been described in section IV and we give in detail their construction here. First, the
$\eta$ coordinates are defined by

$$
\eqalign{ & {\cal O}(2) \quad\quad {\tilde\eta} = z + x \rho - {\bar z} \rho^2 = {\bar z}
(\rho-\rho_+)  (\rho-\rho_{-})
\cr & {\cal O}(4) \quad\quad
\eta = z + v\rho + w\rho^2 -{\bar v}\rho^3 + {\bar z}\rho^4 \cr & \quad\quad\quad\quad\quad = {\bar z}
(\rho -
\alpha)(\rho-\beta)(\rho+{1\over{\bar\alpha}})  (\rho-{1\over{\bar\beta}}) }
\eqn\etacoords
$$ We have two types of integrals and their derivatives with respect to $w$ in eqn.(\use\etacoords).
The four solutions to equations of the type

$$
\eta' \equiv {\tilde\eta}^2 - {\eta\over 4} =0 \eqn\quartic
$$ give the various limits of the integration. Note that this equation satisfies the reality condition

$$ {\bar {\eta(\rho)'}} = \eta'(- {1\over {\bar\rho}}) {\bar\rho}^4 \ , \anoneqn
$$ so that the four roots come in complex pairs. We define the numbers $\delta_\pm$ and
$-{1/{\bar\delta}_\pm}$ to be the solutions to the equations

$$
\left[ {\sqrt{\eta}\over 2} \pm \eta \right] (\delta_\pm) =0 \anoneqn
$$ By the reality condition the $\pm$-solutions come in pairs $\delta_{\pm}$ and $-1/
{\bar\delta}_\pm$.

\bigskip
\noindent{$\underline{\it Mixed}$}
\bigskip

The mixed $O(2)$, $O(4)$ contour integral is given by 

$$ J_{\tilde C} = {1\over 2\pi i} \oint_{\tilde c} d\rho {1\over \rho^2}  ({\tilde\eta} +
\sqrt{\eta})
\ln({\tilde\eta} + \sqrt{\eta}) \ . \eqn\mixed
$$ There are four zeroes in the argument of the logarithm, two on each of the sheets of the square
root. We take the contour ${\tilde C}$ in a figure eight around the two of the zeroes $\delta_+$ and
$-1/{\bar\delta}_+$ on the plus sheet, illustrated as one of the two contours in fig.(3). The integral
in eqn.(\use\mixed) with the square root sign flipped is taken around the points
$\delta_-$ and $-1/{\bar\delta}_-$. The orientation of the contours around the zeroes of $\eta=0$
should be noted. In all cases we begin the analytic continuation in the region where the zeroes of
$\eta=0$ and ${\tilde\eta} \pm \sqrt{\eta}=0$ live on the real axis and are ordered as

$$
\delta_- \leq -1/{\bar\delta}_+ \quad\leq~ \alpha \leq -1/{\bar\beta} \leq  -1/{\bar\alpha} \leq
\beta ~\leq\quad -1/{\bar\delta_-} \leq \delta_+ \ , \eqn\inequality
$$ as drawn in fig.(3).

\bigskip
\noindent{$\underline{{\cal O}(4)~w{\it-Derivative}}$} \bigskip

The $w$-derivative on $F^{12}_{\rm int}$ gives the integral 

$$ I_C = {1\over 2\pi i} \oint_{C} d\rho~ {1\over\sqrt{\eta}} \anoneqn
$$ The contour $C$ is illustrated in fig.(2). The zeroes of the $O(4)$ coordinate satisfy the reality
constraint and are given by $z_i= \left\{ \alpha,~-1/{\bar\beta},~-1/{\bar\alpha},~\beta
\right\}$. 

We begin the analytic continuation of the integral in the region where all $z_i$ are real and ordered
so that $z_i < z_{i+1}$. The contour integral may be deformed into a standard elliptic integral 

$$
\eqalign{ I_C & = - \int_{\alpha}^{-1/{\bar\beta}} d\rho {1\over\sqrt{\eta}} \cr & = -2 ({{\bar\alpha}
{\bar\beta}\over {\bar z}})^{{1\over 2}}  {1\over
\left[(1+\beta{\bar\beta})(1+\alpha{\bar\alpha})\right]^{1\over 2}} {\cal K}(k) \ . }
\anoneqn
$$ The modulus is purely real and given by

$$ k^2 = {(1+{\bar\alpha}\beta)(1+\alpha{\bar\beta})\over  (1+\alpha{\bar\alpha})(1+\beta{\bar\beta})}
\ . \anoneqn
$$

\vskip -.3 cm
\LoadFigure\OFourContour {\baselineskip 13 pt
\noindent\narrower\ninerm The contour for the integral $I_C$.} {\epsfxsize 4.0 truein \epsfysize 1.5
truein}{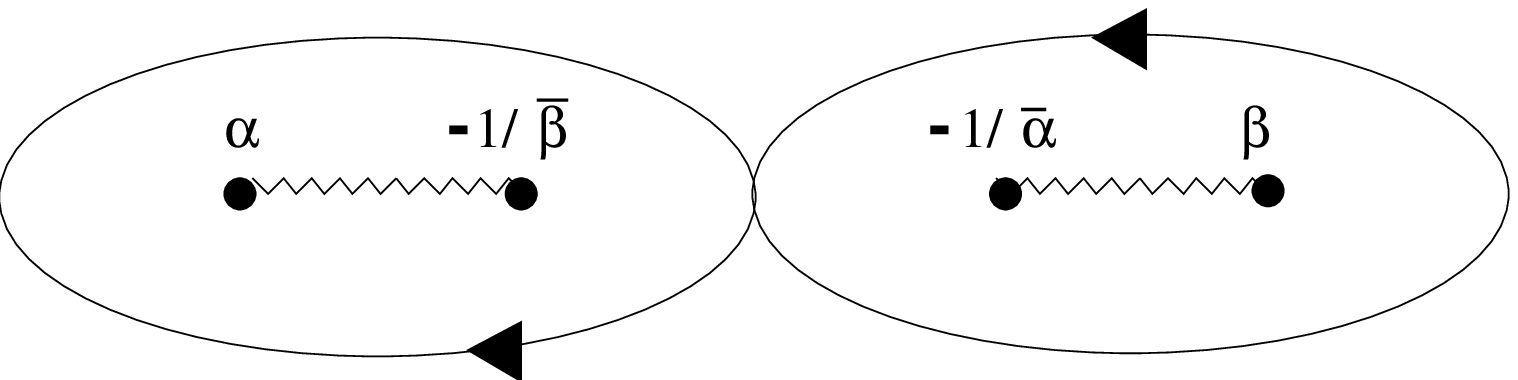}{} \vskip .2in

\bigskip
\noindent{$\underline{{\it Mixed}~w{\it-Derivative}}$} \bigskip

Now we compute the $w$-derivative of $J_{\tilde C}$ in eqn.(\use\mixed). The integral is denoted by
$I_{\tilde C} =
\partial_w J_{\tilde C}$, and the monodromy around the product of the logarthim and square root gives
the integral on the top of the real axis as 

$$
\eqalign{ I_{\tilde C} & = {1\over 4\pi i} ~\bigl( \int_{\delta_+}^{-1/{\bar\delta}_+} + 
\int_{-{\bar\delta}_+}^{\delta_+} \bigr)~ d\rho {1\over\sqrt{\eta}} 
\ln( {\tilde\eta} + \sqrt{\eta}) + {1\over 2} \int_{\delta_+}^{-1/{\bar\delta}_+} d\rho
{1\over\sqrt{\eta}} \cr & = {1\over 2} \int_{-1/{\bar\delta}_+}^{\delta_-} d\rho {1\over\sqrt{\eta}} \
. }
\anoneqn
$$ The contribution from the complement integral on the other sheet of the square-root surface gives
the sum

$$ I_{{\tilde C}_+} + I_{{\tilde C}_-} = {1\over 2} 
\Bigl( \int_{-1/{\bar\delta}_+}^{\delta_-} + 
\int^{-1/{\bar\delta}_-}_{\delta_+} \Bigr) d\rho {1\over\sqrt{\eta}} \ .
\eqn\mixedtotal
$$ The line segments are generically away from any poles which may be encountered in the denominator.
The combination of both integrals has no ambiguity in defining the square root branch cut. Also, the
total integral in eqn.(\use\mixedtotal) is invariant under the change of variables
$\rho \leftrightarrow -1/{\bar\rho}$. 

The terms in eqn.(\use\mixedtotal) lead to four incomplete integrals, two from each ``side'' of the
elliptic cuts, 

$$
\eqalign{ I_{\tilde C} = (-{{\bar\alpha}{\bar\beta}\over{\bar z}})^{1\over 2} 
\left[(1+\alpha{\bar\alpha})(1+\beta{\bar\beta})\right]^{-{1\over 2}} \Bigl( & {\cal
F}[\phi_R(\delta_+),q] - {\cal F}[\phi_R(-1/{\bar\delta}_-),q]
\cr & + {\cal F}[\phi_L(\delta_-),q] - {\cal F}[\phi_L(-1/{\bar\delta}_+),q]
\Bigr) \ . }
\eqn\mixedexplicit
$$ Note that the explicit form is invariant under the complete inversion of the inequality
\use\inequality . The notation is defined as follows. The incomplete integral ${\cal F}[\phi,k]$ of
the first kind is defined by

$$ {\cal F}[\phi,q] \equiv \int^{\sin{\phi}}_0 {dt \over \left[  (1-t^2)(1-q^2 t^2)
\right]^{1\over 2}} \ . \anoneqn
$$ The angles and (real) modulus in this case are 

$$ q^2 = {({\bar\beta}-{\bar\alpha})(\beta-\alpha)\over  (1+\beta{\bar\beta})(1+\alpha{\bar\alpha})}
\anoneqn
$$ and
$$
\sin(\phi^R(x)) = \Bigl[ - {(1+\alpha{\bar\alpha}) (x -\beta) 
\over (\beta-\alpha) (1+{\bar\alpha}x)} \Bigr]^{1\over 2} \quad\quad
\sin(\phi^L(x)) = \Bigl[ - {(1+\beta{\bar\beta}) (\alpha - x) 
\over (\beta-\alpha) (1+{\bar\beta}x)} \Bigr]^{1\over 2} \ . \anoneqn
$$ The net result for the K\"ahler potential and its derivatives will contain many terms of the form
in eqn.(\use\mixedexplicit), each representing the interaction between the pair of aligned monopoles
and one of the orthogonally charged ones.

\vskip -.3 cm
\LoadFigure\OTwoFourContour {\baselineskip 13 pt
\noindent\narrower\ninerm The contour for the integral $J_{\tilde C}$.} {\epsfysize 1.9truein
\epsfxsize 4.5 truein}{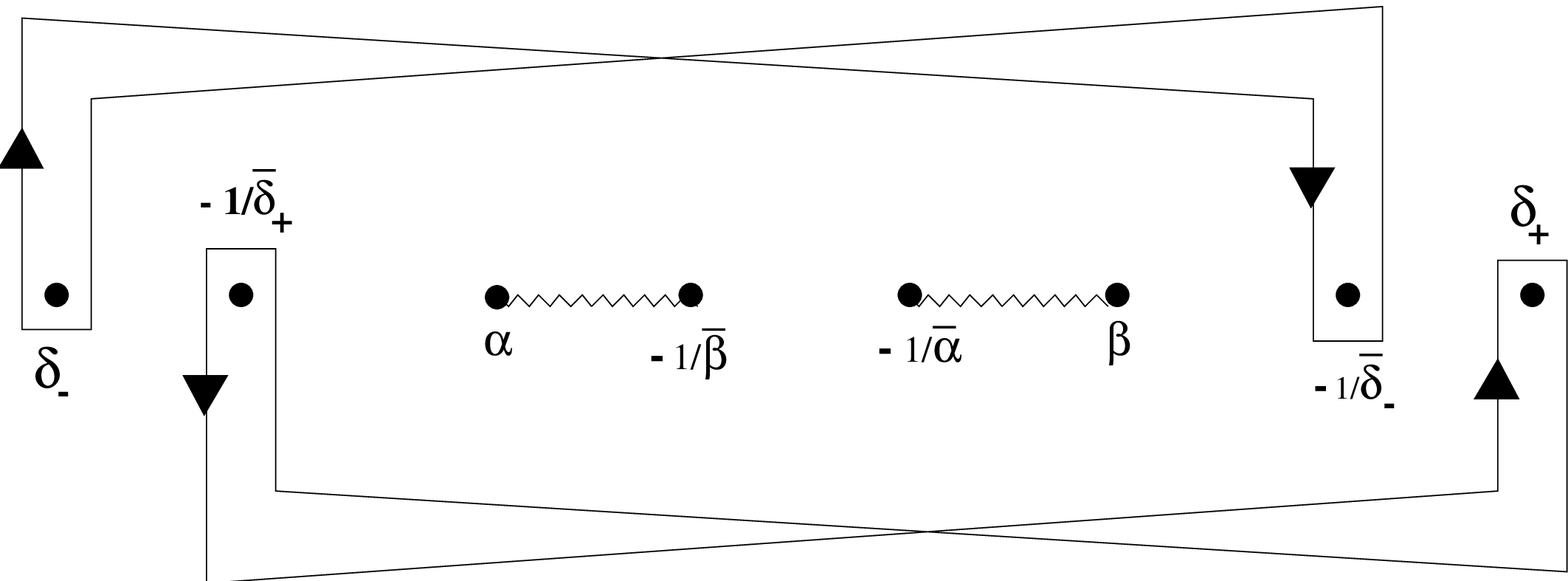}{}

\vfill
\break

\vfill
\baselineskip 14 pt
\listrefs
\end

